\documentclass[11pt,nofootinbib,superscriptaddress]{revtex4}

\usepackage{epsfig,bm,amsmath,amssymb,graphicx,multirow,mathrsfs}

\newcommand{\sdot}{\!\cdot\!}
\newcommand{\SLASH}[1]{/\!\!\! #1}

\begin{document}

\title{The Scalar Meson $\bm {f_0(980)}$ in Heavy-Meson Decays}

\author{B.~El-Bennich} 
\affiliation{Laboratoire de Physique Nucl\'eaire et de Hautes \'Energies (IN2P3--CNRS--Universit\'es Paris 6 et 7), Groupe Th\'eorie, 
                  Universit\'e Pierre et Marie Curie, 4 place Jussieu, 75252 Paris, France}
\affiliation{Physics Division, Argonne National Laboratory, Argonne, Illinois, 60439, USA}
                  
\author{O.~Leitner}      
\affiliation{Laboratoire de Physique Nucl\'eaire et de Hautes \'Energies (IN2P3--CNRS--Universit\'es Paris 6 et 7), Groupe Th\'eorie, 
                  Universit\'e Pierre et Marie Curie, 4 place Jussieu, 75252 Paris, France}    
\affiliation{INFN, Laboratori Nazionali di Frascati, Via E. Fermi 40, I-00044 Frascati, Italy} 
                  
\author{J.-P. Dedonder} 
 \author{B.~Loiseau}
\affiliation{Laboratoire de Physique Nucl\'eaire et de Hautes \'Energies (IN2P3--CNRS--Universit\'es Paris 6 et 7), Groupe Th\'eorie,
                  Universit\'e Pierre et Marie Curie, 4 place Jussieu, 75252 Paris, France}
                    
 \date{\today}

\begin{abstract}
A phenomenological analysis of the scalar meson $f_0(980)$ is performed that relies on the quasi-two body decays $D$ and $D_{s} \to f_0(980)P$, with $P=\pi,\ K$.  
The two-body branching ratios are deduced from experimental 
data on $D$ or $D_s \to \pi\pi\pi,\ \bar K K\pi$ and from the $f_0(980)\to \pi^+\pi^-$ and $f_0(980)\to K^+K^-$ branching fractions. 
Within a covariant quark model, the scalar form factors 
 for the transitions $D$ and $D_s \to f_0(980)$ are computed. 
The weak $D$ decay amplitudes, in which these 
form factors enter, are obtained in the naive factorization approach assuming a $q\bar{q}$ state for the scalar and pseudoscalar mesons. 
They allow to extract  information on the $f_0(980)$ wave function in terms of $u\bar{u},\ d\bar{d}$ and $s\bar{s}$ pairs as well as on 
the mixing angle between the strange and nonstrange components. 
The weak transition form factors are modeled by the one-loop triangular diagram using two different relativistic approaches: covariant light-front dynamics and dispersion relations. 
We use the information found on the $f_0(980)$ structure to evaluate
the scalar and vector form factors
in the transitions $D$ and $D_s \to f_0(980)$, as well as to make 
predictions  for  $B$ and $B_s \to f_0(980)$, for the entire kinematically allowed momentum range of $q^2$.
\end{abstract}


\maketitle

\section{Introduction}\label{section1}

Scalar mesons have been a recurrent topic over the past 30--40 years. 
Whereas the existence of the  $\sigma/f_0(600)$ has been a longstanding open question since the 1960s, the $f_0(980)$ and $a_0(980)$ were firmly 
established in $\pi\pi$ scattering experiments in the 1970's~\cite{Protopopescu:1973sh}. 
The known $0^{++}$ mesons fall into two classes: near and about 1~GeV and in the region $1.3-1.5$~GeV. 
The scalar objects below 1~GeV form an $SU(3)$ flavor nonet~\cite{Leitner:2007pf}. 
This nonet contains two isosinglets, an isotriplet and two strange isodoublets. 
Among these lighter scalars, the isosinglet  $f_0(980)$ and the isotriplet  $a_0(980)$ are rather narrow with their widths $\Gamma$ ranging from 40
 to 100~MeV~\cite{Yao:2006px}. 
Both scalars strongly couple to the $\bar KK$ channel and lie close to its threshold at 987~MeV. 
This closeness alters the shape of  their resonant structure and the description of the $f_0$ and $a_0$ requires a coupled-channel scattering analysis. 
The simple quark model views these scalar mesons as orbitally ($L=1$) excited $\bar qq$ states and has been advocated, for example, by T\"ornqvist and
 Roos~\cite{Tornqvist:1995ay} as well as in Ref.~\cite{Bediaga:2003zh}. 
However, some studies \cite{Black:1998wt} tend to favor four-quark configurations of the scalar mesons, as do coupled-channel analyses~\cite{Krupa:1995fc} or 
potential models of molecular states strongly coupled to $\pi\pi$ and $\bar KK$ channels~\cite{Weinstein:1990gu}. 

The emergence of the $f_0(980)$ as a pole of the $\pi\pi$ amplitude in the $S$ wave~\cite{Kaminski:1998ns} is also well established in three-body decays of
 $B$ mesons~\cite{El-Bennich:2006yi}. 
Recent $\pi\pi$ effective mass range distributions, obtained from an isobar model fit of the
$B\to \pi^+\pi^-K$ and $B\to \bar KK K$  Dalitz plots by the Belle~\cite{Garmash:2005rv,Garmash2005,Abe0509047} 
and BaBar Collaborations~\cite{AubertPRD72,AubertPRD73,Aubert0408}, display distinct peaks about 1~GeV. 
Scalar resonances have also been observed in the charmed three-body decays $D\to \pi\pi\pi,\pi\pi K,\bar KK K$ at CLEO~\cite{Muramatsu:2002jp}, 
FOCUS~\cite{Link:2002iy}, ARGUS~\cite{Albrecht:1993jn}, BaBar~\cite{Aubert:2002yc}, 
E791~\cite{Aitala:2000xt,Aitala:2002kr,Aitala:2000xu}, 
E687~\cite{Frabetti:1995sg,Stenson:2001yg}.
Remarkably, in Ref.~\cite{Aitala:2000xu} an experimental evidence for a light and broad scalar resonance in the $m_{\pi\pi}$ spectrum of the
 $D\to \pi\pi\pi$ decay was found, which may be identified with the $f_0(600)$ and a 
 peak within the $f_0(980)$ mass range is also observed. 
Although a considerable amount of data has been accumulated over the years, it has yet not been possible to elucidate the precise $f_0(980)$ 
quark structure, {\em i.e.\/} whether one deals with a two-quark or rather a four-quark  composite, and thus far there is no consensus on that matter. 
On the other hand, viewing the $f_0(980)$ exclusively as a $\bar qq$ or  ${\bar q}^2 q^2$ state may simply be too naive \cite{Pennington:2007eg}. 
In this context, an interesting proposition to shed light on the constituent composition 
of the $f_0(980)$ was recently made by Maiani, Polosa and Riquer~\cite{Maiani:2007iw}. Their method consists in comparing the ratio of the decay 
rates $D_s^+\to \pi^+(K^+K^-)$ and $D_s^+\to \pi^+(K_S K_S)$. 
This ratio is predicted to be $1/2$ if the $f_0(980)$ is an $I=0$, $\bar qq$ state, whereas the composition $f_0=[sq][\bar s\bar q], q=u,d$, could 
yield a different value owing to possible interference patterns between $I=0$ and $I=1$ amplitudes in the tetraquark picture of these decays. 
For a general overview on scalar mesons, we refer to the Particle Data Group review \cite{Yao:2006px} and references therein.

In the case of $B\to f_0(980)K$ decays, one may advance plausible reasons to limit oneself to the $\bar qq$ picture of the $f_0(980)$. 
Because of the large $B$ mass, the outgoing mesons are virtually massless particles, which prompts to expand the corresponding bound states in terms of Fock states. 
Quark configurations like ${\bar q}^2 q^2$ or ${\bar q}^2 q^2 g$ therefore belong to higher Fock states. 
A handwaving argument by Cheng, Chua and Yang \cite{Cheng:2005nb} suggests that 
the $\bar qq$ component of the energetic $f_0(980)$ may be more important, as two rapid $\bar qq$ pairs are less likely to form a fast moving $f_0(980)$. 
In our models we neglect higher Fock contributions to the $f_0(980)$ bound state. 

In the two-quark model of the light scalar octet below 1~GeV, assuming an ideal $SU(3)$ mixing angle of the octet states, the flavor content of the  $f_0(980)$ is purely strange
($s \bar s$) while that of the $\sigma$ or $f_0(600)$ is purely nonstrange ($u \bar u+ d \bar d$) (see e.g.~\cite{Cheng:2005nb}).
In such a picture the $\sigma$ is the lightest scalar, the $f_0(980)$  the heaviest. 
However, there is compelling experimental evidence,
for instance from $f_0(980) \to \pi\pi$ decays~\cite{Yao:2006px}, that the $f_0(980)$ cannot be made of strange quarks only.  
We therefore introduce in this work some mixing between the strange and nonstrange flavor content. 
Experimental implications on this mixing have been the object of several studies (see e.g.~\cite{Cheng:2002ai,AnisovichEPJA01,AnisovichPAT02,Gokalp05})

In this paper we complete preliminary calculations of $B\to f_0(980)$ and $D\to f_0(980)$ pseudoscalar to scalar ($P\to S$) transition form 
factors~\cite{ElBennich:2006yy,Leitner:2006hu}. Transition form factors are important  for an understanding of the hadronic component of heavy-to-light decay amplitudes and their precise  evaluation is crucial to  a reliable determination of Cabibbo-Kobayashi-Maskawa (CKM) matrix elements. 
The short-distance physics is calculated in perturbative QCD, which comprises radiative vertex corrections to local four-quark operators in the  
operator product expansion~\cite{Buras:1998ra} as well as hard-scattering corrections with the spectator quark that go beyond the leading order~\cite{Beneke:2001ev}. In contrast, the transition form factors are by nature long-distance nonperturbative hadronic matrix elements. 
They provide one ingredient of the factorizable amplitudes of the nonleptonic $B$ decays mentioned above. 
A variety of theoretical approaches to heavy-to-light transition (pseudoscalar to pseudoscalar) amplitudes exist, either using light-cone sum 
rules~\cite{Khodjamirian:2006st}, light-front~\cite{Lu:2007sg} or relativistic quark models~\cite{Ebert:2006nz,Ivanov:2000aj,Faessler:2002ut,Melikhov:2001zv}. 
Most recently, a comprehensive set of $B$ meson heavy-to-light transition form factors, calculated with truncated amplitudes based on Dyson-Schwinger equations 
in QCD, was reported in Ref.~\cite{Ivanov:2007cw}. Whereas the methods of Refs.~\cite{Khodjamirian:2006st,Lu:2007sg} only provide form factors for a small 
domain of timelike momentum transfers $q^2$, those in Refs.~\cite{Ebert:2006nz,Ivanov:2000aj,Faessler:2002ut,Melikhov:2001zv,Ivanov:2007cw} 
give access to the entire range of physical timelike momenta. To our knowledge, $P\to S$ transition form factors have only been evaluated so far with QCD sum rules~\cite{Cheng:2005nb,Aliev:2007uu} at $q^2=0$ four-momentum transfer. A functional extrapolation is required to access all timelike $q^2$ in these studies.

The present work relies on two explicitly covariant formalisms: the covariant light-front dynamics (CLFD) and the dispersion relation (DR) approaches.
Both require two size parameters as well as a mixing angle between the $u\bar{u},\ d\bar{d}$ and $s\bar{s}$ components to specify  the $f_0(980)$ wave function.
 In order to deduce these parameters from experiment, we fit $D$ and $D_s\to f_0(980)P$ branching fractions, where $P$ can be a pion or a kaon. 
Initially, decay amplitudes at tree level in the naive factorization approach are employed and neither annihilation nor penguin topologies are considered. 
Already at tree level, nonleptonic two-body $D$ decays can be reasonably reproduced within this simple factorization since penguin amplitudes are strongly 
CKM suppressed. However, since the charm mass $m_c$ is lighter than the bottom mass by roughly a factor three, nonperturbative contributions of order
 $\Lambda_{\mathrm{QCD}}/m_c$ are more important than in $B$ decay amplitudes. The factorization approach may then be less reliable. In order to study the discrepancy between theoretical and experimental branching fractions, we study the effect of phenomenological annihilation as well as penguin amplitudes. 
The decay amplitudes are proportional to the $D$ and $D_s \to f_0(980)$ transition form factors we are interested in.

The  paper is organized as  follows.  In Sec.~\ref{sec2}, we present the CLFD formalism and give a brief review on the DR approach. 
The scalar $f_0(980)$ bound-state structure is described in Sec.~\ref{sec3}.  In Sec.~\ref{sec4}, we list all physical constraints imposed in our model, namely 
experimental $D$-branching ratios and wave-function normalizations. The electroweak decay amplitudes, the $D$-decay tree topologies and all numerical inputs needed are presented in Sec.~\ref{sec5}.  In Sec.~~\ref{sec6} we introduce the $P\to S$ transition form factors, derived in CLFD and DR approaches. 
Details about the initial pseudoscalar wave functions, the pseudoscalar decay constant in the constituent quark model and the calculation of the $P\to S$ 
transition form factors  are given in Appendices~\ref{appA}, and~\ref{appB}. In Sec.~\ref{sec7}, we discuss the fitting method, give numerical results for the theoretical 
branching ratios and then compare $D_{(s)}\to f_0(980)$ and $B_{(s)}\to f_0(980)$ transition form factors obtained in both relativistic approaches\footnote{Here and in the following, the notation $D_{(s)}$ refers either to the $D$ or to the $D_s$ meson and similarly $B_{(s)}$ refers to the $B$ or to the $B_s$ meson.}.
The final Sec.~\ref{sec8} summarizes our work and some conclusions are drawn.

\section{Two different relativistic formalisms\label{sec2}}
\subsection{Covariant Light-Front Dynamics\label{clfd}}

In CLFD~\cite{Carbonell:1998rj}, the state vector which describes the physical bound 
state is defined on the light-front plane given by $\omega \sdot r = \sigma$, where $\omega$  is an unspecified lightlike  four vector ($ \omega^{2} = 0$)
which  defines the position of the light-front plane and $r$ is a four vector position of the system. CLFD proposes a formulation in which the evolution for a given system is expressed in terms of covariant expressions.  Any four vector describing a phenomenon can be transformed from a system of reference to another by using a unique standard matrix which depends only on  kinematical parameters and on $\omega$. The particle is described by a wave function expressed in terms of Fock components of the state vector  which respects the properties required under any transformation. The meson of mass $M$ will be described as a bound state of two constituent quarks with four momenta $k_1$ and $k_2$. The state vector describing this meson of four-momentum $p$, defined on a light-front plane characterized
 by $\omega$, is given by:
\begin{multline}\label{eq7.8}
{\vert p, \lambda \rangle}_{\omega} = (2 \pi)^{3/2} \int \Phi_{j_{1}\sigma_1j_2
\sigma_2}^{J\lambda}
(k_1,k_2,p,\omega \tau)a_{\sigma_{1}}^{\dagger}({\bf k}_1)a_{\sigma_2}^{\dagger}({\bf k}_2)\vert 0\rangle \\
 \times
\delta^{(4)}(k_1+k_2-p-\omega \tau) \exp(i \tau \sigma)2(\omega \sdot p) {\rm d}\tau 
\frac{{\rm d}^3 k_1}{(2 \pi)^{3/2}\sqrt{2 \varepsilon_{k_1}}} 
\frac{{\rm d}^3 k_2}{(2 \pi)^{3/2}\sqrt{2 \varepsilon_{k_2}}},
\end{multline}
where $\varepsilon_{k_{i}}=\sqrt{{\bf k}_{i}^2+m_{i}^2}$ and ${\bf k}_{i}$ is the 
momentum of the quark $i$ with mass $m_i$.  The parameter $\tau$ is entirely determined by the on-mass shell condition for the individual constituents.
In Eq.~(\ref{eq7.8}) $\lambda$ is the projection of the total angular momentum $J$ of the system on the $z$-axis in the rest frame and $\sigma_i$ is the 
spin projection of the quark $i$ in the corresponding rest systems. We emphasize that the bound state wave function is always an off-energy shell object, 
with $\tau \neq  0$  due to the binding energy, and depends on the light-front orientation. From the delta function ensuring momentum conservation, one gets:
\begin{equation}\label{eq7.9}
\mathcal{P}= p+ \omega \tau = k_{1}+k_{2}.
\end{equation}
To keep track of this conservation law, a momentum, $\omega \tau$, is assigned to 
the spurion but there is no fictitious particle in the physical state vector, (see Fig.~1).
The two-body wave function $\Phi(k_1,k_2,p,\omega \tau)$  in Eq.~(\ref{eq7.8}) can be  parametrized in terms of various sets of variables. 
In order to make a close connection to the nonrelativistic case, it is more convenient to introduce the following  pair of variables \cite{Carbonell:1998rj} defined by
\begin{equation}
\label{kgras}
{\bf k}=L^{-1}(\mathcal{P}){\bf k}_1={\bf k}_1-\frac{\mathcal{\vec P}}{\sqrt{\mathcal{P}^2}}
\Biggl(k_{10}-\frac{{\bf k}_1\cdot\mathcal{{\vec P}}}{\sqrt{\mathcal{P}^2}+\mathcal{P}_0}
\Biggr),
\end{equation}
\begin{equation}
\label{ngras}
{\bf n}=\frac{L^{-1}(\mathcal{P}){\mbox{\boldmath $ \omega$}}}{\vert L^{-1}(\mathcal{P}) \mbox{\boldmath $\omega$} \vert}
=\sqrt{\mathcal{P}^2}\ \frac{L^{-1}(\mathcal{P}){\mbox{\boldmath $\omega$}}}{\omega\cdot p},
\end{equation}
where $\mathcal{P}=k_1+k_2$, and $L^{-1}(\mathcal{P})$ is the Lorentz boost.
The momentum, ${\bf k}$, corresponds, in the center of mass frame where ${\bf k}_1+{\bf k}_2={\bf 0}$, to the usual relative momentum between the two particles.
Note that this choice of variable does not assume that one is restricted to this particular frame.
The unit vector ${\bf n}$ corresponds, in this frame, to the spatial direction of $\omega$.

One introduces  the variables $x$ and the vector $ R_{1}=(R_{0},{\bf R}_{\perp}, {\bf R}_{\|})$ where ${\bf R}_{\perp}, {\bf R}_{\|}$ denotes the perpendicular
 and parallel components to the direction of the light-front:
\begin{equation}
\label{xR1}
x= \frac{\omega \sdot k_{1}}{ \omega \sdot p},\ 
R_{1}=k_{1}-xp.
\end{equation}
Since by construction $R_1 \sdot \omega=0$, and thus $R_1^2=-{\bf R}_{\perp}^2$, the light-front coordinates, 
which one will use in the present work, are then $(x,{\bf R}_{\perp})$. These variables can be expressed in terms of the ones in Eqs.~(\ref{kgras})
and (\ref{ngras}).  All details can be found in Ref.~\cite{Carbonell:1998rj}.

In terms of the variables $(x,\bf{R}_{\perp})$, we have for the relative momentum between two quarks of different masses:
\begin{equation}
\label{grask2}
{\bf k}^2=\frac{\Bigl\{{\bf R}_{\perp}^2+\bigl[(x-1)m_1-x m_2\bigr]^2\Bigr\}\Bigl\{{\bf R}_{\perp}^2+\bigl[(x-1)m_1-x m_2\bigr]^2\Bigr\}}{4x(1-x)\Bigl[{\bf R}_{\perp}^2+(1-x)m_1^2+xm_2^2\Bigr]}.
\end{equation}

\subsection{Dispersion Relation  approach}

The dispersion relation approach, in the context of the relativistic quark model, leads to 
transition amplitudes expressed as relativistic spectral integrals over spectral 
densities of the corresponding Feynman diagrams. 
Here we closely follow the derivation of Melikhov~\cite{Melikhov:2001zv} to calculate the $P\to S$ transition form factors. 
These are given by the double spectral representation over the square of the invariant masses of the initial and final quark-antiquark bound states. 
The spectral functions involve the wave functions of the participating mesons and the double discontinuities of the  corresponding triangle Feynman diagram. 
Use of the Landau-Cutkosky rules allows to calculate these discontinuities and hence the transition form factors in the space-like region $q^2<0$. 
An analytical continuation in $q^2$ gives the form factors in the timelike region $q^2>0$.

As in Sec.~\ref{clfd}, the meson of mass $M$ is a bound state of two constituent quarks of mass $m_1$ and $m_2$ and four-momentum $k_1$ and $k_2$ with
\begin{equation}
\label{sinvariant}
s=(k_1+k_2)^2,\quad k_1^2=m_1^2,\ k_2^2=m_2^2.
\end{equation}
The relativistic bound state corresponds to a pole in the amplitude at $s=M^2$ and  one can define a bound state wave function $\psi(s)$ in 
the vicinity of the pole by
\begin{equation}
\label{phiG}
\psi(s)=\frac{G_v(s)}{s-M^2}.
\end{equation}
The function $G_v(s)$ in Eq.~(\ref{phiG}) represents the vertex of the bound state transition to the constituent quarks.
The constituent-quark rescatterings lead to the normalization condition~\cite{Melikhov:2001zv}
\begin{equation}
\label{normG}
\int_{(m_1+m_2)^2}^\infty \frac{G_v^2(s)\ \rho(s,m_1,m_2)}{\pi(s-M^2)^2}ds\ =1,
\end{equation}
where the spectral density $\rho(s,m_1,m_2)$ for a pseudoscalar meson reads
\begin{equation}
\label{rhops}
\rho_{P}(s,m_1,m_2)=\frac{\lambda^{1/2}(s,m_1^2,m_2^2)}{8\pi s}\ [s-(m_1-m_2)^2]\ 
\theta\!\left(s-(m_1+m_2)^2\right),
\end{equation}
while for a scalar meson one has
\begin{equation}
\label{rhos}
\rho_S(s,m_1,m_2)=\frac{\lambda^{1/2}(s,m_1^2,m_2^2)}{8\pi s}\ [s-(m_1+m_2)^2]\ 
\theta\!\left(s-(m_1+m_2)^2\right).
\end{equation}
In Eqs. (\ref{rhops}) and (\ref{rhos}), $\lambda(s,m_1^2,m_2^2)$ is defined as
\begin{equation}
\label{lambda}
\lambda(s,m_1^2,m_2^2)\equiv (s+m_1^2-m_2^2)^2-4sm_1^2,
\end{equation}
and $\theta(z)$ is the step function, $\theta(z)=1$ for $z>0$ and $\theta(z)=0$ 
for $z<0$.

From Eqs.~(\ref{rhops}), (\ref{rhos}) and (\ref{lambda}) it can be inferred with 
$m_1 = m_2 =m$ that the threshold behaviors of $\rho_{P}(s,m_1,m_2)\propto(s-4m^2)^{1/2}$ and of $\rho_{S}(s,m_1,m_2) \propto (s-4m^2)^{3/2}$ correspond to those of
 an $S$ and of a $P$ wave, respectively. 
Taking into account the intrinsic negative parity of the $\bar qq$ state, it implies the correct behavior under parity transformation of the bound state
 described by the $\bar qq$ state and its associated vertex (see Eq.~(\ref{vertexs})).

\section{Structure of the bound state for a scalar particle \label{sec3}}

Assuming that the $f_0(980)$ scalar meson is made of  components $u\bar{u}, d\bar{d}$ and $s\bar{s}$, one can decompose the total wave function as follows:
\begin{eqnarray}
\label{psif0}
\Psi_{f_0}=\frac{1}{\sqrt{2}} (u\bar{u}+ d\bar{d})\sin \theta_{\mathrm{mix}} +  s\bar{s} \cos \theta_{\mathrm{mix}},
\end{eqnarray} 
or 
\begin{eqnarray}\label{eqwf}
\Psi_{f_0}=\Psi_{f_0^{(n)}} \sin \theta_{\mathrm{mix}} + \Psi_{f_0^{(s)}} \cos \theta_{\mathrm{mix}}
=N_S(\, \phi^{(n)}\sin\theta_{\mathrm{mix}}+\phi^{(s)}\cos\theta_{\mathrm{mix}} \, ),
\end{eqnarray} 
where $\theta_{\mathrm{mix}}$ is the mixing angle between the nonstrange, $\Psi_{f_0^{(n)}}$, and strange, $\Psi_{f_0^{(s)}}$, flavor content of the wave function\footnote{Consequently this implies a strange component for the wave function of the $\sigma$, $\Psi_\sigma =(u \bar u + d \bar d)\cos \theta_{\mathrm{mix}}/\sqrt{2}   - 
s \bar s \sin\theta_{\mathrm{mix}}$. However such a strangeness content does not seem to have an experimental support
 (see for instance Ref.~\cite{Gokalp05}). This certainly points to  a more involved structure  of the $\sigma$ or  $f_0(600)$  than that of a simple 
 $q \bar q$ state.}. In what follows, unless otherwise stated, $m_n$ will denote the up or down quark mass ($m_n=m_u=m_d$), $m_s$ that of the strange quark and $N_S$ is the normalization constant of the full wave function.

\subsection{The scalar particle on the light front}

\begin{figure}
\begin{center}
\includegraphics*[width=0.5\columnwidth]{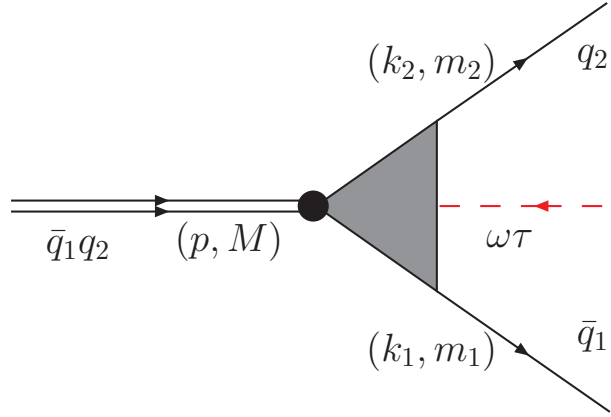}
\end{center}
\caption{Representation of the two-body wave function on the light front.} 
\label{fig1} 
\end{figure}

The explicit covariance of this approach allows to write the general structure of the two-body bound state. 
For a scalar particle (see Fig.~\ref{fig1}) composed of a  quark-antiquark pair of equal mass $m_q$ and four-momenta $k_2$ and $k_1$, we have ($q=n$ or $s$)
\begin{equation}\label{eq58.1}                                                    
\phi^{(q)}=\frac{1}{\sqrt{2}}\bar{u}(k_2) A^{(q)}(x,{\bf R}^2_\perp) v(k_1),
\end{equation}
where  $v(k_{1})$ and  ${\Bar u}(k_{2})$  are the usual antiparticle and particle Dirac spinors, and $A^{(q)}(x,{\bf R}^2_\perp)$ is the  scalar component
 of the wave function.  
Note that the color factor is not included in the wave function Eq.~(\ref{eq58.1}). 
Since the quark masses $m_q$, in each component $A^{(q)}(x,{\bf R}^2_\perp)$, are identical, the corresponding reduced mass is $m_q/2$ and we chose the following 
Gaussian expression:
\begin{eqnarray}\label{defaclfd}
A^{(q)}(x,{\bf R}^2_\perp) =  \exp \Bigl(-16\ \nu_q  {\bf k}^2_q/m_q^2 \Bigr),
\end{eqnarray} 
where $\nu_q$ is a size parameter to be determined  from experimental data and theoretical assumptions while  
the momentum squared, ${\bf k}^2_q$, given in Eq.~(\ref{grask2}) now reduces to 
\begin{equation}
\label{eq:k2}
{\bf k}^2_q=\frac{{\bf R}_{\perp}^{2}+m_q^2(2x-1)^2}{4x(1-x)}.
\end{equation}

\subsection{The scalar particle in the dispersion approach}

The soft constituent-quark structure of the scalar meson is given in this approach by the vertex 
\begin{equation}
\label{vertexs}
\frac{\bar Q^a(-k_2)i Q^a(k_1)}{\sqrt{N_C}}\ G_v(s),
\end{equation}
where $Q^a(-k_2)$ and $Q^a(k_1)$ are the constituent spinor states of color $a$ normalized by the color factor $N_C=3$. 
For a scalar meson made of a quark-antiquark pair of equal mass $m_q$, the wave function $\phi^{(q)}(s)$ of Eq.~(\ref{eqwf})  can be parametrized as 
\begin{equation}
\label{eq:phiS}
\phi^{(q)}(s)=\frac{\pi}{\sqrt{2}}\ \frac{s^{1/4}}{\sqrt{s-4m_q^2}}w_q(k_q),
\end{equation}
where $k_q$ is the modulus of the quark momentum in the center of mass momentum such that
\begin{equation}
\label{eq:4kq2}
4 {\bf k}_q^2=s-4m_q^2.
\end{equation}
The functional form (\ref{eq:phiS}) is so chosen as will be seen later, so as to simplify the normalization condition in Eq.~(\ref{normG}). 
The function $w(k)$ is defined to have the same functional expression as in CLFD:
\begin{equation}
\label{wk}
w_q(k)= \exp\left(-16\ \nu_q{\bf k}_q^2/m_q^2\right),
\end{equation}
and here again the size parameter $\nu_q$ is to be determined from experimental and theoretical considerations.

\section{Physical constraints for a neutral scalar}\label{sec4}

As described in detail previously, the $\bar qq$ bound states are described in both formalisms by vertex functions which are related to Gaussian wave functions. 
These have to be normalized and their phenomenological size parameters determined.  A standard approach chosen in the quark model is to calculate the decay 
constant with the appropriate loop diagram and fix the size parameter that enters the calculation so as to reproduce the experimental value of that constant (see Appendix~\ref{appA}). In this work, this is done for the pseudoscalar $D$- and $B$-meson wave functions. However, the lack of knowledge of the experimental 
$f_0(980)$ decay constant makes it difficult to proceed similarly for the scalar meson. Furthermore, the mixing angle $\theta_{\mathrm{mix}}$ is not known
{ \em a priori}. We therefore resort to a different parametrization prescription by making use of $D$ decay branching ratios which contain the $f_0(980)$ 
 in the final state. In this section, we discuss the constraints on the scalar wave functions given by  the normalization and the experimental data set chosen to determine the mixing angle $\theta_{\mathrm{mix}}$ as well as the various size parameters in both formalisms.

\subsection{Normalization in CLFD}

According to the spirit of the constituent quark model, the state vector is decomposed into Fock components, and only the two-body component is retained. Since the state vector is normalized as
\begin{equation}\label{eq8.16}
\langle p^{\prime},\lambda^{\prime} | p, \lambda \rangle = 2 p_{0} \delta^{(3)}\ (\bf p -
\bf p^{\prime})\ \delta^{\lambda^{\prime} \lambda},
\end{equation} 
it gives for a zero total angular momentum state the following normalization 
condition~\cite{Carbonell:1998rj}:
\begin{equation}
1 = \int_{(x,\tilde\theta,{\bf R}_{\perp})} 
D(x,\tilde\theta,{\bf R}_{\perp})
\sum_{\lambda_{1} \lambda_{2}} \Phi_{\lambda_{1} \lambda_{2}}^{(q)} 
\Phi_{\lambda_{1} \lambda_{2}}^{(q)\star},
\end{equation}
where $D(x,\tilde\theta,{\bf R}_{\perp})$, is the invariant phase space element given by:
\begin{equation}
\label{eq9.19}
D(x,\tilde\theta,{\bf R}_{\perp})
=\frac{1}{(2\pi)^3}\frac{{\rm d}^2{\bf R}_{\perp}{\rm d} x}{2x(1-x)}.
\end{equation}
Using the condition of normalization for the Dirac spinors, 
$\sum_\lambda\ u_a^\lambda(k)\bar u_b^\lambda(k)=(\SLASH k+m)_{ab}$ and 
$\sum_\lambda\ v_a^\lambda(k)\bar v_b^\lambda(k)=(\SLASH k-m)_{ab}$, we sum over all spin and color states and get for a  $q\bar q$ component:
\begin{align}
\label{sumlambda12}
\sum_{\rm color}\sum_{\lambda_1,\lambda_2} \Phi_{\lambda_1,\lambda_2}^{(q)}\  \Phi_{\lambda_1,\lambda_2}^{(q)\dagger}& =  \dfrac{1}{2}\sum_{\rm color}
\sum_{\lambda_1,\lambda_2}\bar u^{\lambda_2}(k_2)A^{(q)}v^{\lambda_1}(k_1)\bar v^{\lambda_1}(k_1)A^{(q)}u^{\lambda_2}(k_2), \nonumber\\
 & =  \dfrac{1}{2} \sum_{a,b,c,d} \sum_{\lambda_1,\lambda_2} u_d^{\lambda_2}(k_2)\bar u_a^{\lambda_2}(k_2)(A^{(q)})_{ab}
v_b^{\lambda_1}(k_1)\bar v_c^{\lambda_1}(k_1)(A^{(q)})_{cd},\nonumber\\
 & =   \dfrac{1}{2}\sum_{a,b,c,d} (\SLASH k_2+m)_{da}(A^{(q)})_{ab} (\SLASH k_1-m)_{bc}(A^{(q)})_{cd},\nonumber\\
 & =   \dfrac{1}{2}\mathrm{Tr}  \left[(\SLASH k_2+m)  (\SLASH k_1-m) (A^{(q)})^2 \right],
\end{align}
where $A^{(q)}$ is given by Eq.~(\ref{defaclfd}). The result is similar for both the $n\bar n$ and $s\bar s$ components.
There is no mixing term between the two components.
With the scalar wave function written in Eq.~(\ref{eqwf}), the normalization condition is therefore:
\begin{equation}
\label{eq8.19}
1= N_S^2  \int_{(x,\tilde\theta,{\bf R}_{\perp})} 
\Bigg [ 
\frac{{\bf k}_n^2}{4m_n^2} \bigl(A^{(n)}(x,{\bf R}^2_\perp)\bigr)^2 \sin^2 \theta_{\mathrm{mix}}
+\frac{{\bf k}_s^2}{4m_s^2}\bigl(A^{(s)}(x,{\bf R}^2_\perp)\bigr)^2 \cos^2 \theta_{\mathrm{mix}}
 \Bigg ]D(x,\tilde\theta,{\bf R}_{\perp}),
\end{equation}
with ${\bf k}_q$ given by Eq.~(\ref{eq:k2}).

\subsection{Normalization in the Dispersion Relation approach}

In the DR approach, for the scalar meson, the wave function [see in Eqs.~(\ref{phiG}),~(\ref{eqwf}) and~(\ref{eq:phiS})] 
is normalized according to Eq.~(\ref{normG}). Taking into account  the quark-content assumption of the $f_0(980)$
introduced in Eq.~(\ref{eqwf}) and making use of the form for 
$\phi^{(n)}(s)$ or $\phi^{(s)}(s)$ given by Eq.~(\ref{eq:phiS}), the normalization condition for  $\Psi_{f_0}$ reads
\begin{equation}\label{normdr}
 1 =  N_S^2 \int_0^\infty k^2 \big [w_n^2(k)\sin^2\theta_{\mathrm{mix}} + w_s^2(k)\cos^2\theta_{\mathrm{mix}} \big ] dk,
\end{equation}
since the cross contributions vanish because of the orthogonality of the flavor states. In Eq.~(\ref{normdr})  $w_n(k)$ is the nonstrange 
Gaussian component of $\phi^{(n)}(s)$ whereas $w_s(k)$ 
is the strange one
of $\phi^{(s)}(s)$, which implies two different size parameters $\nu_n$ and  $\nu_s$. 
The form~(\ref{wk}) for $\omega_q(k)$ leads to the normalization
\begin{equation}
 N_S = \frac{2}{\pi^{1/4}} \Biggl[\left ( \frac{m_n^2}{32\nu_n} \right )^{\!\!\frac{3}{2}}\!\sin^2\theta_{\mathrm{mix}}+
\left ( \frac{m_s^2}{32\nu_s} \right )^{\!\!\frac{3}{2}}\!\cos^2\theta_{\mathrm{mix}} 
  \Biggr]^{\!\!-\frac{1}{2}}.
\label{dispnorm}
\end{equation}

%
\subsection{$\boldsymbol{D}$ meson branching ratios\label{expdata}}

The wave function for scalar meson $f_0(980)$ -- denoted hereafter for simplicity $f_0$ -- is constrained by the experimental branching ratios for the 
channels $D^+ \to f_0 \pi^+, D^0 \to f_0 \bar{K}^0, D^+ \to f_0 K^+, D_s^+ \to f_0 
\pi^+, D_s^+ \to f_0 K^+$ and $D^0 \to f_0 \pi^0$. The experimental ratios are provided by different collaborations, E791~\cite{Aitala:2000xt,Aitala:2002kr,Aitala:2000xu}, ARGUS~\cite{Albrecht:1993jn}, 
CLEO~\cite{Muramatsu:2002jp,CroninHennessy:2005sy,Eidelman:2004wy}, BABAR~\cite{Aubert:2002yc}, FOCUS~\cite{Link:2002iy} and E687~\cite{Frabetti:1995sg,Stenson:2001yg}:

\begin{align}\label{eq:26a}
&E791~[22,23]: \;\;  &\mathcal{B}(D^+ \to f_0 \pi^+) \times \mathcal{B}(f_0 \to \pi^+ \pi^-)=
&(1.9 \pm 0.5)\times 10^{-4}, 
\\ \nonumber \\ \label{eq:26b}
&ARGUS~[19]: \;\; & \mathcal{B}(D^0 \to f_0 \bar{K}^0)\times \mathcal{B}(f_0 \to \pi^+ \pi^-)
=&(3.2 \pm 0.9)\times10^{-3}, 
\nonumber \\ 
&CLEO~[17]: \;\;  &\mathcal{B}(D^0 \to f_0 \bar{K}^0) \times\mathcal{B}(f_0 \to \pi^+ \pi^-)
=&(2.5^{+0.8}_{-0.5})\times10^{-3},
\nonumber \\
&BABAR~[20]: \;\; & \mathcal{B}(D^0 \to f_0 \bar{K}^0)\times \mathcal{B}(f_0 \to K^+ K^-)
=&(1.2 \pm 0.9)\times10^{-3}, 
\\ \nonumber\label{eq:26c}
\end{align}
\begin{align}
&FOCUS~[18]: \;\; &\mathcal{B}(D^+ \to f_0 K^+)  \times\mathcal{B}(f_0 \to K^+ K^-)
=&(3.84 \pm 0.92)\times10^{-5}, 
\nonumber \\ 
&FOCUS~[18]: \;\; & \mathcal{B}(D^+ \to f_0 K^+) \times\mathcal{B}(f_0 \to \pi^+ \pi^-)
=&(6.12 \pm 3.65)\times10^{-5},  
\\ \nonumber \\ \label{eq:26d}
&E687~[24,25]: \;\;  & \mathcal{B}(D_s^+ \to f_0 \pi^+) \times\mathcal{B}(f_0 \to K^+ K^-)
=&(4.9 \pm 2.3)\times10^{-3}, 
\nonumber \\
&E791~[21]: \;\;  & \mathcal{B}(D_s^+ \to f_0 \pi^+)\times \mathcal{B}(f_0 \to \pi^+ \pi^-)
=&(5.7 \pm 1.7)\times10^{-3}, 
\nonumber \\
&FOCUS~[18]: \;\; &  \mathcal{B}(D_s^+ \to f_0 \pi^+)\times \mathcal{B}(f_0 \to \pi^+ \pi^-)
=&(9.5 \pm 2.7)\times10^{-3}, 
\nonumber \\
&FOCUS~[18,29]: \;\; &  \mathcal{B}(D_s^+ \to f_0 \pi^+) \times\mathcal{B}(f_0 \to K^+ K^-)
=&(7.0 \pm 1.9)\times10^{-3}, 
 \\ \nonumber \label{eq:26e}
\end{align}
\begin{align}
&FOCUS~[18]: \;\; &  \mathcal{B}(D_s^+ \to f_0 K^+) \times\mathcal{B}(f_0 \to K^+ K^-)
=&(2.8 \pm 1.3)\times10^{-4},
\\ \nonumber \\ \label{eq:26f}
& CLEO~[46,47]: \;\; & \mathcal{B}(D^0 \to f_0 \pi^0) \simeq  \mathcal{B}(D^0 \to \pi^+ 
\pi^- \pi^0)&\times 
\mathcal{F}(D^0 \to f_0 \pi^0)=  \nonumber\\  
 & &(1.1 \pm 0.4)\times10^{-2} \times  (1.0 \pm 0.8)\times10^{-4}, 
\end{align}
In Eq.~(\ref{eq:26f}) $\mathcal{F}(D^0 \to f_0 \pi^0)$ represents the fit 
fraction of the $(D^0 \to f_0 \pi^0)$ decay~\cite{CroninHennessy:2005sy}\footnote{The value used for $\mathcal{B}(D^0 \to \pi^+ \pi^- \pi^0)$ is taken from Ref.~\cite{Eidelman:2004wy} in order to be 
consistent with that of Ref.~\cite{CroninHennessy:2005sy}. The more recent and precise value from Ref.~\cite{Yao:2006px} does not modify our conclusions.}.
The $f_0(980)$ width is dominated by the $f_0$ decay into $\pi\pi$ and $K\bar K$.
Combining their partial wave analysis~\cite{Ablikim:2005} of $\chi_{c_0}\to\pi^+\pi^-K^+K^-$ with their study~\cite{Ablikim:2004} of
 $\chi_{c_0}\to f_0f_0\to\pi^+\pi^-\pi^+\pi^-$ the BES Collaboration~\cite{Ablikim:2005} has determined the following ratio between the partial widths of the $f_0$
\begin{equation}
\label{eq:R}
R=\frac{\Gamma(f_0\to\pi\pi)}{\Gamma(f_0\to\pi\pi)+\Gamma(f_0\to K\bar K)}=0.75_{-0.13}^{+0.11}.
\end{equation}
Applying isospin relations, one finds the following branching fractions
\begin{equation}
\label{eq:brf0pipi}
\mathcal{B}(f_0\to\pi^+\pi^-)=\frac{2R}{3}=0.50_{-0.09}^{+0.07},
\end{equation}
\begin{equation}
\label{eq:brf0KK}
\mathcal{B}(f_0\to K^+K^-)=\frac{1-R}{2}=0.125_{-0.065}^{+0.055}.
\end{equation}
The two-body branching ratios $\mathcal{B}(D\to f_0P)$ entering Eqs.~(\ref{eq:26a}) to (\ref{eq:26f}) are then deduced from the branching fractions
 $\mathcal{B}(f_0\to \pi^+\pi^-)$ and
$\mathcal{B}(f_0\to K^+K^-)$ given in Eqs.~(\ref{eq:brf0pipi}) and (\ref{eq:brf0KK}).
It is worth emphasizing that the results are strongly dependent on these branching fractions. 
Note that their experimental uncertainties are large.

\section{Electroweak amplitude\label{sec5}}

\noindent In any phenomenological treatment of the weak decays of hadrons, the 
starting point is the weak effective Hamiltonian, which is obtained by 
integrating out the heavy fields  from the standard model Lagrangian and reads
\begin{equation}
\label{eq1}
{\cal H}_{eff}^{\bigtriangleup C=1}=\frac {G_{F}}{\sqrt 2}
\sum_i V_{\mbox{\tiny CKM}} C_{i} (\mu)  O_i (\mu) + h.c.,
\end{equation}
where $G_{F}$ is the Fermi constant, $V_{\mbox{\tiny CKM}}$ contains products 
of the CKM matrix element, $C_{i}(\mu)$ are the Wilson coefficients, $O_i(\mu)$ are the operators entering the operator product expansion and $\mu$ 
represents the renormalization scale.
In the present case, since we only take into account  tree  operators, the matrix 
elements of the  Hamiltonian (\ref{eq1}) read,
\begin{eqnarray}\label{eq2}
\langle M_1 M_2|{\cal H}_{eff}^{\bigtriangleup C=1}| D \rangle= 
\frac {G_{F}}{\sqrt 2} \sum_q \Biggl[  V_{cq}V_{uq}^{*} 
\sum_{i=1}^{2} C_{i}(\mu)\langle M_1 M_2 |O_{i}^{q}|D
 \rangle(\mu)  \Biggr]\ + \ h.c.,
\end{eqnarray}
where $q=d \; {\rm or} \; s$ according to the transition $c \to d$ or $c \to s$.  The scale $\mu$ is chosen to be of order $m_c$ for $D$ decays.
The amplitudes $\langle M_1 M_2 |O_i^q| D \rangle(\mu)$ are hadronic matrix elements,  $M_1$ and  $M_2$ denote a pseudoscalar and a scalar meson in the final state.  In Eq.~(\ref{eq2}) the notation $\langle M_1 M_2 |O_{i}^{q}|D\rangle(\mu)$ reflects the fact that the hadronic matrix elements also depend on the renormalization scale $\mu$.  They describe the transition amplitude between initial and final states at scales 
lower than $\mu$ and give rise to the main uncertainties in the calculation, as they involve the nonperturbative regime of QCD. 
The operator product expansion divides the calculation of the amplitude $A(D \rightarrow M_1M_2)\propto C_{i}(\mu) \langle M_1M_2| O_{i} | D \rangle (\mu)$ 
into two distinct physical regimes. One regime deals with hard or short-distance physics, represented by the Wilson coefficients $C_{i}(\mu)$ and
calculated perturbatively, the other concerns soft or long-distance physics. The operators $O_{i}(\mu)$ can be understood as local operators which govern
 a given decay, reproducing the weak interaction of quarks in a point-like approximation. The Wilson coefficients $C_{i}(\mu)$~\cite{Buras:1998ra} contain the physical contributions from scales higher than $\mu$. Since  QCD has the property of asymptotic freedom, they can be calculated in perturbation theory and include contributions from all heavy particles with $m>\mu$, such as the top  and beauty quarks, the $W^{\pm}$ bosons, and the charged Higgs boson. 
The dependence of the hadronic matrix elements and of the $C_{i}(\mu)$ on $\mu$ must cancel in the final decay amplitude which is a physical observable and 
thus scale independent. 

Working at tree level within the  factorization formalism one obtains the following decay amplitudes: 
\begin{equation}
    \mathcal{A}(D \to f_0 P)=\left\{\,\,
   \begin{array}{l}
      \frac{G_F}{2} V_{cd} V_{ud}^{*}\ (m_{D^{+}}^2-m_{f_0}^2) a_1 f_{\pi} 
         \mathcal{F}_0^{D^+ \to f_0^{(n)}}(m^2_{\pi^+})\sin \theta_{\mathrm{mix}}\,
      \;\;{\rm for} \;\; D^+ \to f_0 \pi^+,  \\[0.4cm] 
     \frac{G_F}{2} V_{cd} V_{us}^{*}\ (m_{D^{+}}^2-m_{f_0}^2) a_1 f_{K} 
\mathcal{F}_0^{D^+ \to f_0^{(n)}}(m^2_{K^+})\sin \theta_{\mathrm{mix}}\,  
       \;\; {\rm for} \;\; D^+ \to f_0 K^+, \\[0.4cm]
      \frac{G_F}{2} V_{cs} V_{ud}^{*}\ (m_{D^{0}}^2-m_{f_0}^2)  a_2 f_{K} 
         \mathcal{F}_0^{D^0 \to f_0^{(n)}}(m^2_{K^0})\sin \theta_{\mathrm{mix}}\,
       \;\;{\rm for} \;\; D^0 \to f_0 \bar{K}^0,  \\[0.4cm]
    \frac{G_F}{\sqrt{2}} V_{cs} V_{ud}^{*}\ (m_{D^{+}_s}^2-
m_{f_0}^2) a_1 f_{\pi} \mathcal{F}_0^{D^+_s \to f_0^{(s)}}(m^2_{\pi^+})\cos \theta_{\mathrm{mix}}\,
     \;\; {\rm for} \;\; D^+_s \to f_0 \pi^+, \\[0.4cm]
    \frac{G_F}{\sqrt{2}} V_{cs} V_{us}^{*}\ (m_{D^{+}_s}^2-
m_{f_0}^2) a_1 f_{K}  \mathcal{F}_0^{D^+_s \to f_0^{(s)}}(m^2_{K^+})\cos \theta_{\mathrm{mix}}\ 
    \; {\rm for} \; D^+_s \to f_0 K^+, \\[0.4cm] 
     \frac{G_F}{2\sqrt{2}} V_{cd} V_{ud}^{*}\ (m_{D^{0}}^2-
m_{f_0}^2) a_2 f_{\pi}  \mathcal{F}_0^{D^0 \to f_0^{(n)}}(m^2_{\pi^0})\sin \theta_{\mathrm{mix}}\,  
       \;\; {\rm for} \;\; D^0 \to f_0 \pi^0,
     \end{array}\right.
     \label{decayamplitudes}
\end{equation}
where $a_i(m_c)$ is written as $a_i$ for simplicity.
In Eq.~(\ref{decayamplitudes}), $f_\pi$ and $f_K$ are the pion and kaon decay constants and
\begin{equation}
\label{eq:a1a2}
a_1(m_c)=C_1(m_c)+\frac{C_2(m_c)}{N_C},\qquad a_2(m_c)=C_2(m_c)+\frac{C_1(m_c)}{N_C},
\end{equation}
where $N_C=3$.
The flavor content $u$ or $s$ of the $D$ and $f_0$ has been written explicitly in the scalar transition form factors $\mathcal{F}_0^{D\to f_0}(m_P^2)$.
With these factorized decay amplitudes, we can compute the decay rates using the 
following expression~\cite{Yao:2006px},
\begin{equation}\label{eq5.1}
\Gamma(D \rightarrow f_0 P)=\frac{1}{8\pi}\frac{\vert {\bf p}\vert }{m_D^2} 
|\mathcal{A}(D \to f_0 P)|^2,
\end{equation}
where $\vert {\bf p}\vert $ is the modulus of the c.m. momentum of the decay particles  defined as
\begin{equation}\label{eq4.16}
|{\bf p}|=\frac{ \sqrt{ [m_{D}^{2}-(m_{P}+m_{f_0})^{2}][m_{D}^{2}-(m_{P}-m_{f_0})^{2}]}}{2m_{D}}.
\end{equation}
Finally, one defines the branching ratio $\mathcal{B}$ as the ratio between the decay rate $\Gamma(D \rightarrow f_0 P)$ and the total decay width $\Gamma_D$:
\begin{eqnarray}\label{br}
\mathcal{B}=\frac{\Gamma(D \rightarrow f_0 P)}{\Gamma_D}.
\end{eqnarray}

\subsection{Numerical inputs}

\subsubsection{Values of CKM matrix elements and Wilson coefficients}\label{sec5.3.1}
%
In the present numerical calculations,  the CKM matrix elements,  expressed  in terms  of the Wolfenstein parameters $A$ and $\lambda$~\cite{Wolfenstein:1983yz, 
Wolfenstein:1964ks} rely on the latest values extracted from charmless semileptonic $D$ decays~\cite{Yao:2006px}:
\begin{equation}\label{eq5.10}
\lambda=0.2257, \qquad  A=0.814.
\end{equation}
The Wilson coefficients, at the mass scale $\mu=m_c$, are $C_1(m_c)=1.3777$ and 
$C_2(m_c)=-0.6941$ (see Ref.~\cite{Buras:1998ra}) from which we infer
\begin{equation}
  a_1(m_c) =  1.1463, \qquad a_2(m_c)= -0.2349.
\end{equation}

\subsubsection{Quark masses}\label{sec5.3.2}
%
%
We use the subsequent standard constituent quark masses  to calculate the transition 
form factors within the quark model approximation. 
\begin{align}\label{eq5.11}
m_{u}=m_{d}& = 0.350 \;{\rm GeV}, & m_{c}& = 1.620 \;{\rm GeV}, \nonumber\\
m_{s}& = 0.510 \;{\rm GeV}, &  m_{b}& = 4.920 \;{\rm GeV}.
\end{align}
For meson masses, the following values~\cite{Yao:2006px} are used:
\begin{align}\label{eq5.12}
m_{B^{\pm}}& = 5.279 \; {\rm GeV},  & m_{B_{s}}&   = 5.369 \; {\rm GeV}, \nonumber\\
m_{D^{\pm}}& = 1.869   \; {\rm GeV},  & m_{D_{s}}&   = 1.968 \; {\rm GeV}, \nonumber\\
m_{D^{0}}& = 1.864     \; {\rm GeV},    & m_{K^{\pm}}& = 0.493 \;{\rm GeV},  \nonumber\\  
m_{K^{0}}&   = 0.497   \;{\rm GeV},   & m_{f_{0}}& = 0.980  \;{\rm GeV}, \nonumber\\
m_{\pi^{\pm}}& = 0.139 \;{\rm GeV}, & m_{\pi^{0}} & = 0.135 \;{\rm GeV}.
\end{align}
The pseudoscalar decay constants $f_P$ are defined as usual by
\begin{align}\label{eq5.13}
\langle P(p_1) | \bar{q}_{1} \gamma^{\mu} \gamma^{5} q_{2}| 0 \rangle & =  -i f_P p_1^{\mu},
\end{align}
$p_1^{\mu}$ being the momentum of  the pseudoscalar meson and the numerical values we used are,
\begin{align}\label{eq5.14}
f_{\pi} = 132 \; {\rm MeV}, \;\;   & f_{K} = 160 \; {\rm MeV}, \;\; f_D = 222 \; {\rm MeV},  \nonumber\\
f_{D_s} = 274 \; {\rm MeV},  \;\;  & f_B = 180 \; {\rm MeV}, \;\;  f_{B_s}  = 259 \; {\rm MeV}.
\end{align}
The Fermi constant is $G_{F}= 1.166\ 391 \times 10^{-5}  {\rm GeV}^{-2}$~\cite{Yao:2006px} and the world average $D$ life-time values:
\begin{eqnarray}
\tau_{D^{0}}  = 0.410 \pm 0.001\; {\rm ps}, \;\; \tau_{D^{\pm}}  = 1.040 \pm 0.007 \; {\rm ps}, \;\; 
\tau_{D_{s}^{\pm}}  = 0.490 \pm 0.001 \; {\rm ps}.
\end{eqnarray}
yield the total $D$ decay widths $\Gamma_D=1/\tau_D$.
%

\section{Weak decay form factors for $\boldsymbol{P \rightarrow S}$ transitions \label{sec6}}
%
\subsection{Standard form factor notation}
%
The decays of $b$ and $c$ quarks are given by the weak current $J^{\mu}_{b(c)}$ 
(even though only the  ${\Bar q } \gamma^{\mu}\gamma^5 q_{b(c)}$ term is relevant in our case),
\begin{equation}\label{eq9.1}
J^{\mu}_{b(c)} = {\Bar q} \gamma^{\mu} (1- \gamma^5)q_{b(c)},
\end{equation}
where $q$ is a light $u,d$ or $s$  quark. As usual, one can define the physical 
amplitude for a semi-leptonic decay $X \rightarrow Y l {\nu}_{l}$ by the expression
\begin{equation}\label{eq9.2}
{\mathcal M} = \frac{G_{F}V_{ij}}{\sqrt{2}} \langle S  | J^{\mu}| P \rangle 
J^{\mbox{\tiny lep}}_{\mu},
\end{equation}
where $J^{\mbox{\tiny lep}}_{\mu}$ is the leptonic current.  In Eq.~(\ref{eq9.2}), 
$\langle S | J^{\mu}| P \rangle$ is the hadronic matrix element including the 
weak current as defined previously.
Introducing the total four-momentum $K=P_1+P_2$ and the four-momentum transfer $q=P_1-P_2$ where $P_1$ is the four-momentum of the pseudoscalar meson and $P_2$ that of the scalar meson in the final state, the hadronic matrix element can be decomposed as:
\begin{equation}
\label{eq9.3}
\langle S(P_{2}) | J^{\mu}| P(P_{1}) \rangle = K^\mu f_+(q^2)+q^\mu f_{-}(q^{2}), 
\end{equation}
where $f_{+}(q^{2})$ and $f_{-}(q^{2})$ are the transition form factors and $P_{1}$ and $P_{2}$  are respectively the four-momentum 
related to the initial and final particle states of the hadronic current.  Introducing then the scalar $\mathcal{F}_{0}(q^{2})$ and vector $\mathcal{F}_{1}(q^{2})$ form factors, 
the amplitude can be expressed as
\begin{equation}
\label{eq9.4}
\langle S(P_{2}) | J^{\mu}| P(P_{1}) \rangle =  \mathcal{F}_{1}(q^{2}) 
\Bigg[ 
K^\mu-\frac{K\cdot q}{q^2}\ q^\mu
\Bigg]
 + \mathcal{F}_{0}(q^{2}) \Bigg[\frac{K\cdot q}{q^2}\ q^\mu\Bigg], 
\end{equation}
since $K\cdot q=M_1^2-M_2^2$, and $M_1$ and $M_2$ being the  masses of the initial and final meson.
It is straightforward to derive the relationship between the two sets of form factors. 
One obtains
\begin{align}\label{eq9.5}
\mathcal{F}_{1}(q^{2})& = f_{+}(q^{2}), \\ \label{eq9.5b}
\mathcal{F}_{0}(q^{2})& = f_{+}(q^{2}) + \frac{q^{2}}{K\cdot q}f_{-}(q^{2}).  
\end{align}
Note  that at $q^2=0$, $\mathcal{F}_{1}(0)=\mathcal{F}_{0}(0)=f_{+}(0)$.

%
\subsection{CLFD formalism}\label{sec9.1.2}
\begin{figure}
\begin{center}
\includegraphics*[angle=0,width=0.6\columnwidth]{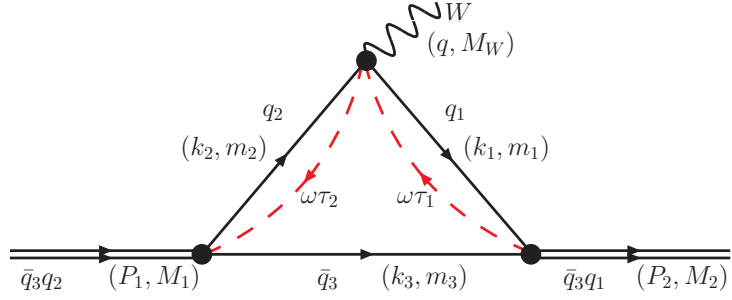}
\end{center}
\caption{The triangle diagram (leading contribution) and momentum flow in the weak-hadronic $P\to S$ transition amplitude of the CLFD approach. In the present case: $m_2 > m_1$.}
\label{triangleCLFD}
\end{figure}

%
In the  covariant light-front dynamics formalism, the exact transition amplitude 
does not depend on the light front orientation but in any approximate computation the dependence  is explicit. However one can 
parametrize this dependence since the formalism 
is covariant. Hence, the approximate amplitude expressed in CLFD is given by the 
following hadronic matrix,
\begin{equation}
\label{eq9.7}
\langle S(P_{2}) | J^{\mu}| P(P_{1}) \rangle^{CLFD} = K^\mu f_+(q^2)+q^\mu f_{-}(q^{2})
     + \omega^{\mu} B(q^{2}), 
\end{equation}
where $B(q^{2})$ is a nonphysical form factor which has to be zero in any exact calculation. The last term represents the explicit dependence of the 
amplitude on the light front orientation $\omega$ with $\omega^2=0$. In order to extract the physical form factor $f{_{\pm}}(q^{2})$, without any dependence
on $\omega$, from the amplitude $\langle S(P_{2}) | J^{\mu}| P(P_{1}) \rangle^{CLFD}$,  we will proceed as  follow. 
First, we calculate the scalar products $\mathcal{X}, \mathcal{Y}$ and $ \mathcal{Z}$ which are defined by,
\begin{equation}
\label{eq9.8}
\mathcal{X} =K_\mu\cdot  \langle S(P_{2}) | J^{\mu}| P(P_{1}) \rangle^{CLFD} =  
K^2f_+(q^2)+K\cdot qf_-(q^2)+ K\cdot\omega B(q^2),
\end{equation}
\begin{equation}
\label{eq9.9}
\mathcal{Y} = q_\mu \cdot \langle S(P_{2}) | J^{\mu}| P(P_{1}) \rangle^{CLFD}  = 
K\cdot q f_+(q^2)+q^2f_-(q^2)+ q\cdot\omega B(q^2),
\end{equation}
and finally,
\begin{equation}
\label{eq9.10}
\omega \sdot P_{1} \; \mathcal{Z} = \omega_{\mu}  \sdot \langle P_{2} | J^{\mu}| P_{1} 
\rangle^{CLFD}=  
K\cdot\omega f_+(q^2)+q\cdot\omega f_-(q^2).
\end{equation}
We define a variable $y$ as the ratio between the scalar product of $\omega \sdot P_{2}$ and $\omega \sdot P_{1}$,
\begin{equation}\label{eq9.11}
y = \frac{\omega \sdot P_{2}}{\omega \sdot P_{1}}= \frac{M_{2}^{2} + P_{1} \sdot P_{2}}{M_{1}^{2}+
 P_{1} \sdot P_{2}},
 \;\; {\rm with} \;\; P_{1} \sdot P_{2} = \frac{1}{2}(M_{1}^{2}+M_{2}^{2}-q^{2}). 
\end{equation}
Since $P_1=(K+q)/2$ and $P_2=(K-q)/2$, we may also write
\begin{equation}
\label{eq:y}
y = \frac{\omega \sdot P_{2}}{\omega \sdot P_{1}}=
\frac{\omega\cdot(K-q)}{\omega\cdot(K+q)}=
\frac{4M_2^2+K^2-q^2}{4M_1^2+K^2-q^2},
\end{equation}
with $\omega \cdot K=(1+y)\omega P_1$ and $\omega \cdot q=(1-y)\omega P_1$. For $q^{2} > 0$, it is convenient to restrict ourselves to the plane defined by 
$\boldsymbol{\omega} \cdot \bf{q} = \bf{0}$.  This condition is allowed  in the system of reference where 
$\bf{P_{1}} + \bf{P_{2}} = \bf{0}$ with ${ P_{1}}_{0} - {P_{2}}_{0} \neq {0}$.  From the scalar products $\mathcal{X}, \mathcal{Y}$ and $\mathcal{Z}$ we can 
isolate the form factors $f_{\pm}(q^{2})$ from  $B(q^{2})$. Then, one gets the expressions for the  form factors $f_{\pm}(q^{2})$:
\begin{equation}
\label{eq9.12}
f_{\pm}(q^{2})= \Omega \;\Psi_{\pm} (y,q^{2},\mathcal{X},\mathcal{Y},\mathcal{Z}),
\end{equation}
where $\Omega$ is identical for both form factors $f_{\pm}(q^{2})$ and can be written as,
\begin{equation}\label{eq9.13}
\Omega  = \frac{1}{4\Bigl[\bigl(M_1^2 (y-1) + q^2  \bigr)y - (y-1) M_2^2    \Bigr]}=\frac{1}{ \bigl[ (y-1) K + (y+1) q\bigr]^2},
\end{equation}
where the functions $\Psi_{\pm} (y,q^{2},\mathcal{X},\mathcal{Y},\mathcal{Z})$ are:
\begin{align}\label{eq9.14}
\Psi_{-} (y,q^{2},\mathcal{X},\mathcal{Y},\mathcal{Z}) = &
(y^{2}-1)\mathcal{X} +(y+1)^{2}\mathcal{Y} \;
+ \Bigr[(1-3 y) M_1^2 - (y-3) M_2^2+ (y-1)q^2    \Bigl]
 \mathcal{Z}, \nonumber \\
\Psi_{+} (y,q^{2},\mathcal{X},\mathcal{Y},\mathcal{Z}) = &
(y-1)^{2}\mathcal{X}+(y^{2}-1) \mathcal{Y}  
+ \Bigl[ (y-1) M_1^2 - (y-1) M_2^2 + (y+1) q^2  \Bigl] \mathcal{Z},
\end{align}
or in terms of the variables $K$ and $q$,
\begin{align}\label{eq9.14b}
\Psi_{-} (y,q^{2},\mathcal{X},\mathcal{Y},\mathcal{Z}) = &
(y^{2}-1)\mathcal{X}+(y+1)^{2}\mathcal{Y}  + \Bigr[(1-y) K^2 - (1+y) K \sdot q\Bigl]
 \mathcal{Z}, \nonumber \\
\Psi_{+} (y,q^{2},\mathcal{X},\mathcal{Y},\mathcal{Z}) = &
(y-1)^{2}\mathcal{X} +(y^{2}-1) \mathcal{Y}  
+ \Bigl[(y-1) K \sdot q + q^{2} (y+1)\Bigl] \mathcal{Z}.
\end{align}
The second step is to express the amplitude $\langle S(P_{2}) | J^{\mu} | P(P_{1})  \rangle^{CLFD}$ without using the form factors $f_{\pm}(q^{2})$. 
The leading contribution to the transition amplitude $\langle S(P_{2}) | J^{\mu} | P(P_{1}) \rangle^{CLFD}$ is given by the diagram shown 
in Fig.~\ref{triangleCLFD}. 
By using the CLFD rules, one can derive the matrix elements from the diagram (Fig.~\ref{triangleCLFD})  and  one has, 
\begin{multline}\label{eq9.15}
\langle S(P_{2}) | J^{\mu} | P(P_{1}) \rangle^{CLFD}_{g} = \\
 \int_{(x,\tilde\theta,{\bf R}_{\perp})} D(x, \tilde\theta,{\bf R}_{\perp})
{\rm Tr} \biggl[ -  \frac{1}{\sqrt{2}} A_S(x^{'},{\bf R}_{\perp}^{'2})(m_1+ \SLASH k_{1})
 i\gamma^{\mu}\gamma^5
(m_{2} 
+ \SLASH k_{2})  \\
\times \frac{1}{\sqrt{2}} A_P^{(qq')}(x,{\bf R}_{\perp}^2)  (m_{3}- \SLASH k_{3}) \biggr]  \frac{1}{1-x^{\prime}},
\end{multline}
where $A_P^{(qq')}(x,{\bf R}_{\perp}^2)$ and $A_S(x^{'},{\bf R}_{\perp}^{'2})$ are the pseudoscalar and scalar wave functions defined in Eq.~(\ref{pseudoeq}) 
and Eq.~(\ref{defaclfd}) respectively. 
Note  that  $x$ and $x^{\prime}$ are the fraction of the momentum carried by a quark $q_{3}$ (the spectator quark) as given by:
\begin{equation}\label{eq9.18}
x= \frac{\omega \sdot k_{3}}{\omega \sdot P_1}, \quad {\rm and}\quad x^{\prime} 
= \frac{\omega \sdot k_{3}}
{\omega \sdot P_2}, \;\; {\rm then} \;\; y=\frac{x}{x'},
\end{equation}
and one also has ${\bf R}_{\perp}^{'}={\bf R}_{\perp}- x' \bf{q}$. Now, one can replace the hadronic matrix element $\langle S(P_{2}) | J^{\mu} 
| P(P_{1}) \rangle^{CLFD}$, which appears  in the scalar products  ${\mathcal X, Y, Z}$  (Eqs.~(\ref{eq9.8},~\ref{eq9.9},~\ref{eq9.10})), 
by the hadronic matrix elements $\langle S(P_{2}) | J^{\mu} | P(P_{1}) \rangle^{CLFD}_{g}$ calculated by applying the CLFD diagrammatic rules 
and given in Eq.~(\ref{eq9.15}).  Hence, by using Eq.~(\ref{eq9.12}) we are  able to compute the form factors
 $f_{\pm}(q^{2})$ as a function of $q^{2}$ and this over the whole  available  four momentum range $0<q^{2}< q^{2}_\mathrm{max}$.

\subsection{Dispersion relation approach}
\begin{figure}
\begin{center}
\includegraphics*[angle=0,width=0.6\columnwidth]{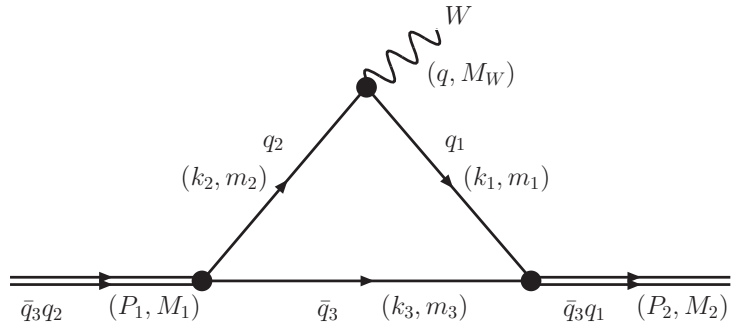}
\end{center}
\caption{Same as in Fig.~\ref{triangleCLFD} but for the DR approach.}
\label{triangleDR}
\end{figure}

%
The pseudoscalar to scalar transition amplitude is calculated from the triangular Feynman diagram shown in Fig.~\ref{triangleDR}, 
where also the kinematical variables are displayed. For the evaluation of the space-like transition form factor $(q^2<0)$ the internal 
constituent quarks are put on-mass shell. Moreover the external momenta are considered off-shell with
\begin{equation}
\label{qoffshell}
\tilde P_1^2=s_1,\qquad  \tilde P_2^2=s_2,\qquad
(\tilde P_1-\tilde P_2)^2=q^2.
\end{equation}
To derive the transition amplitude (\ref{eq9.3}) we need the constituent quark matrix element of the weak axial current which we write
\begin{equation}
\label{f21}
\langle Q_1^{a}(k_1)\vert \bar q_1(0)(-i\gamma^\mu\gamma^5)q_2(0)\vert Q_2^{a}(k_2)
\rangle=-if_{21}(q^2)\bar Q_1^{a}(k_1)\gamma^\mu\gamma^5Q_2(k_2).
\end{equation}
The function $f_{21}(q^2)$ is the constituent quark transition form factor. Since no formal derivation of the quark model from QCD exists,
it is unknown. In the following we make the assumption $f_{21}\simeq 1$ and drop the factor altogether owing to the fact that constituent 
quarks behave very much like bare Dirac particles~\cite{Weinberg:1990xm}.

In the DR approach the transition form factors $f_\pm(q^2)$ of Eq.~(\ref{eq9.3}) are expressed through the double spectral representations:
\begin{equation}
\label{fplusmoins}
f_\pm(q^2)=\int\frac{ds_2\ G_{v_2}(s_2)}
{\pi(s_2-M_2^2)} \int\frac{ds_1\ G_{v_1}(s_1)}{\pi(s_1-M_1^2)}\  \Delta_\pm(s_1,s_2,q^2;m_1,m_2,m_3).
\end{equation}
The functions $\Delta_\pm(s_1,s_2,q^2;m_1,m_2,m_3)$ in the above equation are the double spectral densities of the triangle Feynman of 
Fig.~\ref{triangleDR} in the $P_1^2$- and $P_2^2$- channels. They can be obtained~\cite{Melikhov:2001zv} from the following equation
\begin{eqnarray}
\label{doubles}
\lefteqn{(\tilde P_1+\tilde P_2)^\mu\Delta_+(s_1,s_2,q^2;m_1,m_2,m_3) + 
(\tilde P_1-\tilde P_2)^\mu\Delta_-(s_1,s_2,q^2;m_1,m_2,m_3)} \nonumber\\
 & = &\displaystyle\frac{1}{8\pi}\int d^4k_1d^4k_2d^4k_3\delta(k_1^2-m_1^2)
\delta(k_2^2-m_2^2) \delta(k_3^2-m_3^2)
 \delta(\tilde P_1-k_2-k_3)\delta(\tilde P_2-k_3-k_1) \nonumber \\ 
&\times & \mbox{Tr}\left[-(\SLASH k_{1}+m_1)\gamma^\mu\gamma^5(\SLASH k_{2}+m_2)i\gamma^5
 (m_3-\SLASH k_{3})i\right],
\end{eqnarray}
where $m_2 > m_1$. 
Explicit expressions for $\Delta_\pm(s_1,s_2,q^2;m_1,m_2,m_3)$ are given in Appendix~\ref{appB}. An 
analytical continuation in $q^2$ allows us to write the transition 
form factors for $q^2<(m_2-m_1)^2$ as
\begin{eqnarray}
\label{f+-full}
 f_\pm(q^2) & = & \displaystyle\int\limits_{(m_1+m_3)^2}^\infty
    \frac{ds_2\ G_{v_2}(s_2)}{\pi(s_2-M_2^2)}\       
    \int\limits_{s_1^-(s_2,q^2)}^{s_1^+(s_2,q^2)}\ 
    \frac{ds_1\ G_{v_1}(s_1)}{16\pi(s_1-M_1^2)}\ 
    \frac{B_\pm(s_1,s_2,q^2)}{\lambda^{3/2}(s_1,s_2,q^2)}  \hspace*{1.2cm} \nonumber \\
 & + & 2 \theta(q^2)\displaystyle\int\limits_{s_2^0(q^2)}^\infty
    \frac{ds_2\ G_{v_2}(s_2)}{\pi(s_2-M_2^2)}\ 
    \int\limits_{s_1^R(s_2,q^2)}^{s_1^-(s_2,q^2)}\ 
    \frac{ds_1}{16\pi(s_1-s_1^R)^{3/2}} \nonumber \\
 & \times &  \displaystyle\left[
  \frac{G_{v_1}(s_1)\ B_\pm(s_1,s_2,q^2)}{(s_1-s_1^L)^{3/2}(s_1-M_1^2)}
  -\frac{G_{v_1}(s_1^R)\ B_\pm(s_1^R,s_2,q^2)}{(s_1^R-s_1^L)^{3/2}(s_1^R-M_1^2)}
   \right].
\end{eqnarray}
The functions $s_1^L(s_2,q^2)=(\sqrt{s_2}-\sqrt{q^2})^2$ and $s_1^R(s_2,q^2)
=(\sqrt{s_2}+\sqrt{q^2})^2$ are the roots of
$\lambda(s_1,s_2,q^2)=(s_1+s_2-q^2)^2-4s_2s_1$. The expressions for 
$B_\pm(s_1,s_2,q^2)$ are given in Appendix~\ref{appB}
along with the integration limits $s_1^-(s_2,q^2)$, $s_1^+(s_2,q^2)$ and $s_2^0(q^2)$. 

 We note that although the diagrams for $D\to f_0(980)$ and $\pi \to f_0(980)$ are very similar in the calculation of their spin trace, the main 
 difference is that  no kinematical  factor is involved at the  triangle apex where the interaction vertex $\gamma_\mu (1-\gamma_5)$ in the
 case of the heavy meson decays stems from the weak interaction. In the $\pi \to f_0(980)$ form factor all constituent masses are identical, while in 
 the present case there are two mass scales, namely the charm and a light or strange quark. It can be seen, in Appendix~\ref{appB}, 
 that the expressions for the functions $B_{\pm}(s_1,s_2,q^2)$ which enter the spectral densities vanish identically for $m_1=m_2=m_3$. 
 Indeed, in Ref.~\cite{Anisovich:1998bd}, which relies on the method developped in Refs.~\cite{Anisovich:1995sq,Anisovich:1996hh}, the transition
 amplitude for $\pi \to f_0(980)$ calculated on the light front was shown to vanish as $t\simeq  q^2_\perp \to 0$. This is in contradiction with 
 experimental findings in $\pi^- p \to \pi^0\pi^0 n$  reactions. We can make the parallel for the behavior of our transition amplitude in the limit 
 $q_\perp^2\to 0$  and confirm the vanishing of 
 our form factor for $t=q^2\to 0$ if all internal quark masses are equal. 
 Thus, had we calculated the transition to $f_0(980)$ from a pion, we would obtain  $f_\pm(q^2)=0$ for $q^2\to 0$. We ascribe this discrepancy to our simplified $\bar qq$ picture of the $f_0(980)$ whereas other contributions, likely from pion and kaon clouds, may modify the form factors in particular at low momentum transfer.
  For $m_2\neq m_1=m_3$, however, we deduce from the expressions of our dispersive representation that the integrands in Eq.~(\ref{f+-full}) do not vanish for $q^2 \to 0$, nor do the integrals as confirmed by our numerical calculations.

As mentioned before, the form factor in the region $0<q^2<(m_2-m_1)^2$ can 
be obtained by analytic continuation of the expression in 
Eq.~(\ref{fplusmoins}) for $q^2<0$. In this space-like region, the function 
$\Delta(s_1,s_2,q^2;m_1,m_2,m_3)$ in Eq.~(\ref{delta}) has no square-root cuts 
related to the zeros of $\lambda^{1/2}(s_1,s_2,q^2)$ (they lie on the unphysical
 sheet) and both form factors are given by just the first 
term in Eq.~(\ref{f+-full}). Note that the vertex functions $G_v(s)$ are not 
singular for $s>(m_2+m_3)^2$ and $s>(m_1+m_3)^2$ 
and that $B_\pm(s_1,s_2,q^2)$ are polynomials. Thus, the analytic properties 
of the form factors are determined by the sole behavior 
of the function $\lambda^{1/2}(s_1,s_2,q^2)$ for positive $q^2$. One may study
 the structure of the singularities of the integrand 
in the complex $s_1$ plane for a fixed real value of $s_2> (m_1+m_3)^2$, which
 implies external $s_2$ integration and internal 
$s_1$ integration (interchanging the integration order leads to an equivalent 
integration contour). At $q^2>0$, the square-root cut
endpoint $s_1^R$  moves onto the physical sheet through the interval from $s_1^-$ 
to $s_1^+$ to the left of $s_1^-$. This occurs for 
a value of $s_2>s_2^0(q^2)$, where $s_2^0(q^2)$ is obtained as the solution to 
the equation $s_1^R(s_2,q^2)=s_1^-(s_2,q^2)$ 
and given in Appendix~\ref{appB}. The integration contour of $s_1$ in the complex
 plane must  be deformed so it encompasses 
the points $s_1^R$ and $s_1^+$. It therefore contains two integration segments, 
one being the normal part from $s_1^-$ to 
$s_1^+$ and the other the anomalous part from $s_1^R$ to  $s_1^-$. The double
 spectral density for this anomalous part, on 
the other hand, is obtained from the discontinuity of the function 
$\lambda^{1/2}(s_1,s_2,q^2)$ which can be written as 
$\lambda^{1/2}(s_1,s_2,q^2)=\sqrt{(s_1-s_1^L)(s_1-s_1^R)}$, bearing in mind 
that the branch point $s_1^L$ lies on the unphysical sheet. 
Hence, one has to calculate the discontinuity of $1/(s_1-s_1^R)^{1/2}$ which is
 just twice the the function itself \cite{Lucha:2006vc}. 
This explains the integration limits and the factor two in front of the second 
integral in Eq.~(\ref{f+-full}). The subtraction 
term in the third line of Eq.~(\ref{f+-full}) stems from the function 
$1/\lambda(s_1,s_2,q^2)$ that enters the complete expression 
for $\Delta_\pm(s_1,s_2,q^2;m_1,m_2,m_3)$ in Eq.~(\ref{deltaplusmoins}) and which 
is singular in the lower integration limit $s_1^R$.
It was shown in Ref.~\cite{Ball:1991bs} that an accurate application of the Cauchy 
theorem yields this subtraction term.

\section{Numerical results}\label{sec7}

\subsection{The fit procedure} 
 
As we have discussed in the preceding sections our final aim is to predict  form factors for $B_{(s)} \to f_0$ transitions. To achieve this goal, we first
have to acquire a good knowledge of the $f_0$ wave function.  This will be done through the  evaluation of theoretical branching ratios [Eq.~(\ref{br})] 
for $D_{(s)} \to f_0$ transitions, which implies the calculation of form factors that rely on the $f_0$ wave function, as can be seen in Eqs.~(\ref{eq9.15}) 
for CLFD and~(\ref{f+-full}) for DR. Since on the one hand meson masses and  decay constants are measured, and on the other hand constituent 
quark masses as well as Wilson coefficients are known from theoretical considerations and given in Sec.~\ref{sec5}, the evaluation of the branching ratios
depends only on the $f_0$ wave function parameters:  two size parameters $\nu_n$, $\nu_s$ and the mixing angle $\theta_{\mathrm{mix}}$.  
The overall normalization $N_S$ in Eq.~(\ref{eqwf}) is fixed by means of Eq.~(\ref{eq8.19}) for CLFD and Eq.~(\ref{dispnorm}) for DR. Once the $f_0$ 
wave function parameters are given, the form factors $\mathcal{F}_0^{D_{(s)} \to f_0}(q^2)$ and hence the  branching ratios can be determined. These parameters 
will thus be constrained, via a least-square $\chi^2$ fit\footnote{The routine MINUIT~\cite{James:1975dr} has been used to  minimize the  $\chi^2$ in 
this work.}, by  the experimental branching ratios given in Eqs.~(\ref{eq:26a}) to (\ref{eq:26f}). Note that there are two equivalent solutions 
for $\theta_{\mathrm{mix}}$, as the mixing angle enters quadratically into the decay rate formula Eq.~(\ref{eq5.1}).  
As an additional physical constraint, we choose to impose the relation 
\begin{equation} 
   \nu_s = \frac{m_s}{m_u}\,\nu_n,  
\label{sizecorr} 
\end{equation} 
between the strange and nonstrange components of the $f_0(980)$ wave function. This forces the strange component to  be wider in momentum space, 
the size parameter $\nu_s$ being divided by $m_s^2$ in the Gaussian wave functions given in Eq.~(\ref{defaclfd}) or Eq.~(\ref{wk}), 
assuming that $|\bar ss\rangle$ is more tightly bound and compact in configuration space.  This effectively reduced parametrization proves to be 
decisively more stable, while not spoiling the fit. 

We will see that this first simple approach, a two parameter fit attempting to reproduce all data listed in Eqs.~(\ref{eq:26a}) to (\ref{eq:26f}), to 
which we will refer to as fit 1,  provides, partly because of the large experimental errors, a fair agreement with the data though not entirely satisfactory. 
Indeed, so far we {\em a priori\/} miss relevant physics in these decays such as corrections to simple tree-order topologies. We must include 
higher-order and power suppressed contributions in the appropriate channels.  We here consider penguin and annihilation topologies which we
now discuss in turn.
 
In the decays discussed in Sec.~\ref{expdata}, penguin topologies only contribute to the $D^+\to f_0\pi^+$, $D^+_s\to f_0 K^+$ and $D^0 \to f_0\pi^0$  
amplitudes.  The magnitude of the CKM matrix elements~\cite{Yao:2006px} implies that for charmed penguins the penguin contributions can be and are usually 
discarded since $V_{cd}V^*_{ud} \simeq V_{cs}V^*_{us}$ is three orders of magnitude larger than $V_{cb}V_{ub}^*$. Nonetheless, in order 
to try to even more constrain  the scalar mixing angle, we have inserted phenomenological penguin amplitudes where they are operative.  
	 
We have parametrized these contributions by a universal amplitude so that we have only modified the linear combination of Wilson  coefficients, $a_i$:  
\begin{equation} 
 a_i  \ \Longrightarrow \ a_i+X_p(\rho_p,\delta_p) \;\; {\rm with} \;\;  X_p(\rho_p,\delta_p) = \rho_p \exp (i\delta_p ), 
 \label{penguin} 
\end{equation} 
which leads for the amplitude $\mathcal{A}(D^+\to f_0\pi^+)$ to the substitution 
 \begin{multline} 
  \mathcal{A}(D^+\to f_0\pi^+) \ \Longrightarrow \ \mathcal{A}(D^+\to f_0\pi^+) \\ +\frac{G_F}{2} V_{cd} V_{ud}^{*}\ (m_{D^{+}}^2-m_{f_0}^2)  f_{\pi} 
         \mathcal{F}_0^{D_u^+  \to f_0^u}(m^2_{\pi^+})\sin \theta_{\mathrm{mix}}\ X_p(\rho_p,\delta_p),
\end{multline} 
 and similarly for the other two channels with the same $X_p(\rho_p,\delta_p)=\rho_p \exp (i\delta_p )$. 
 
As has been argued in  Ref.~\cite{Cheng:2002wu,Cheng:2002ai}, weak annihilation amplitudes are not negligible for the decays $D \to PP, SP$ and  
are comparable to the tree amplitudes. This occurs because these annihilation amplitudes, denoted in the literature by  
$W$ exchange  or $W$ annihilation  topologies, can receive contributions from long-distance final-state  
interactions. At the hadronic level, the quark rescattering is manifest in $s$ channel resonances and the $W$-exchange 
 topologies receive contributions from, for example, the $0^-$ resonance $K(1830)$~\cite{Cheng:2002wu}.  
Thus, we introduce a phenomenological annihilation term, $X_a(\rho_a,\delta_a)$, in the $D^0 \to f_0(980) \bar K^0$  
decay channel such that 
\begin{equation} 
  \mathcal{A}(D^0 \to f_0(980) \bar K^0) \ \Longrightarrow \ \mathcal{A}(D^0 \to f_0(980) \bar K^0) 
  + G_F\ X_a(\rho_a,\delta_a)\,\frac{\sin \theta_{\mathrm{mix}}}{2}, 
\label{annihil} 
\end{equation} 
with $X_a(\rho_a,\delta_a) = \rho_a \exp (i \delta_a)$.
The modulus $\rho_a$ and phase $\delta_a$ are free parameters, the natural scale of $\rho_a$ is in principle given by the decay  
constants $f_D^0,\ f_K^0$ and $f_{f_0}$.  We stress that neither the contribution from penguin nor from annihilation amplitudes will allow to resolve 
the ambiguity on the mixing angle $\theta_{\mathrm{mix}}$.

In the following we introduce the effective transition form factors which, in the nonstrange sector read, 
\begin{equation} 
  {F}_{0,1}^{P\to f_0}(q^2) =  \mathcal{F}_{0,1}^{P\to f_0}(q^2) \dfrac{\sin \theta_{\mathrm{mix}}}{\sqrt{2}}
\label{effecffn} 
\end{equation}
and in the strange one
\begin{equation} 
  {F}_{0,1}^{P\to f_0}(q^2) =  \mathcal{F}_{0,1}^{P\to f_0}(q^2) \cos \theta_{\mathrm{mix}}.
\label{effecffs} 
\end{equation} 

In principle, the six parameters should be fit to the branching ratios listed in Eqs.~(\ref{eq:26a}) to (\ref{eq:26f}). It turns out, as expected from 
the arguments given above, that in both approaches, CLFD and DR, the contributions of the penguin amplitudes are vanishingly small and do 
not lead to any improvement of the fit while the mixing angle maximally changes by $1^\circ$. In fact, the phase of the penguin amplitude 
is nearly zero and the modulus is very small. We conclude that we may just ignore its contribution. We will therefore refer from now on 
to fit 2 as a four parameter fit which includes solely the annihilation amplitudes as correction to the tree level. 

Before discussing in details the results of  our calculations, we wish to point out the large experimental errors that appear in the constraining data. 
There are furthermore inconsistencies in these data as can be  seen for instance in the FOCUS experiment~\cite{Link:2002iy}, for the $D_s^+ \to f_0 \pi^+$ 
channel. Here, we observe a discrepancy in the  decay magnitude between the channels where the $f_0(980)$ decays into a two-pion or two-kaon 
pair as well as in their errors. Partly, this may be ascribed to the use of the different branching fractions $\mathcal{B}(f_0 \to \pi^+ \pi^-)$ and  
$\mathcal{B}(f_0 \to K^+ K^-)$. 

Considering the theoretical ratio  $R_1=\mathcal{A}(D^+ \to f_0 \pi^+)/\mathcal{A}(D^+ \to f_0 K^+)$ and the corresponding one for the $D^+_s$ meson,
 $R_2=\mathcal{A}(D_s^+ \to f_0 \pi^+)/\mathcal{A}(D_s^+ \to f_0 K^+)$, one observes that they are equivalent when  working at the tree level approximation for
the decay amplitude if one assumes that $\mathcal{F}_0^{D\to f_0}(q^2)$ has roughly the same value for $q^2=m_{\pi}^2$ and $m_{K}^2$
and similarly for $\mathcal{F}_0^{D_s\to f_0}(q^2)$. Using Eq.~(\ref{decayamplitudes}) for the decay amplitudes of these channels, the ratios $R_1$ and $R_2$  
are proportional to the same CKM matrix elements, $V_{ud}^*$ and $V_{us}^*$, and to the pion and kaon decay constants; they are of the order of 4.

Experimentally, though, this order of magnitude is strongly violated when data from FOCUS ($\mathcal{BR}(D^+ \to f_0 K^+)=(3.07 \pm 1.65) \times 10^{-4}$),
from E687 ($\mathcal{BR}(D_s^+ \to f_0 \pi^+)= (3.92 \pm 2.63) \times 10^{-2}$), as well as from FOCUS 
($\mathcal{BR}(D_s^+ \to f_0 \pi^+)=(5.60 \pm 3.08) \times 10^{-2}$) are used. These data appear to be incompatible with the other data.  Hence, we shall study two cases in the four parameter minimization space, one with 12 data, referred to as fit 2a, 
the other with 9 consistent data referred to as fit 2b.

\subsection{The $f_0$ wave function} 
 
Table~\ref{12b2pbr} which corresponds to fit 1 (with 2 parameters and 12 branching ratios) shows that the factorization model at the tree level order 
allows for a fair representation of the data with reasonably well defined parameters $\nu_n$ and $\theta_{\mathrm{mix}}$ given in Table~\ref{12b2ppa}. 
An obvious discrepancy occurs for the $D_0 \to f_0 \bar{K}^0$ channel, the apparent agreement with the BABAR data~\cite{Aubert:2002yc} being only 
due to the very large experimental error. 

\begin{figure}
\begin{center}
\includegraphics*[angle=0,width=0.78\columnwidth]{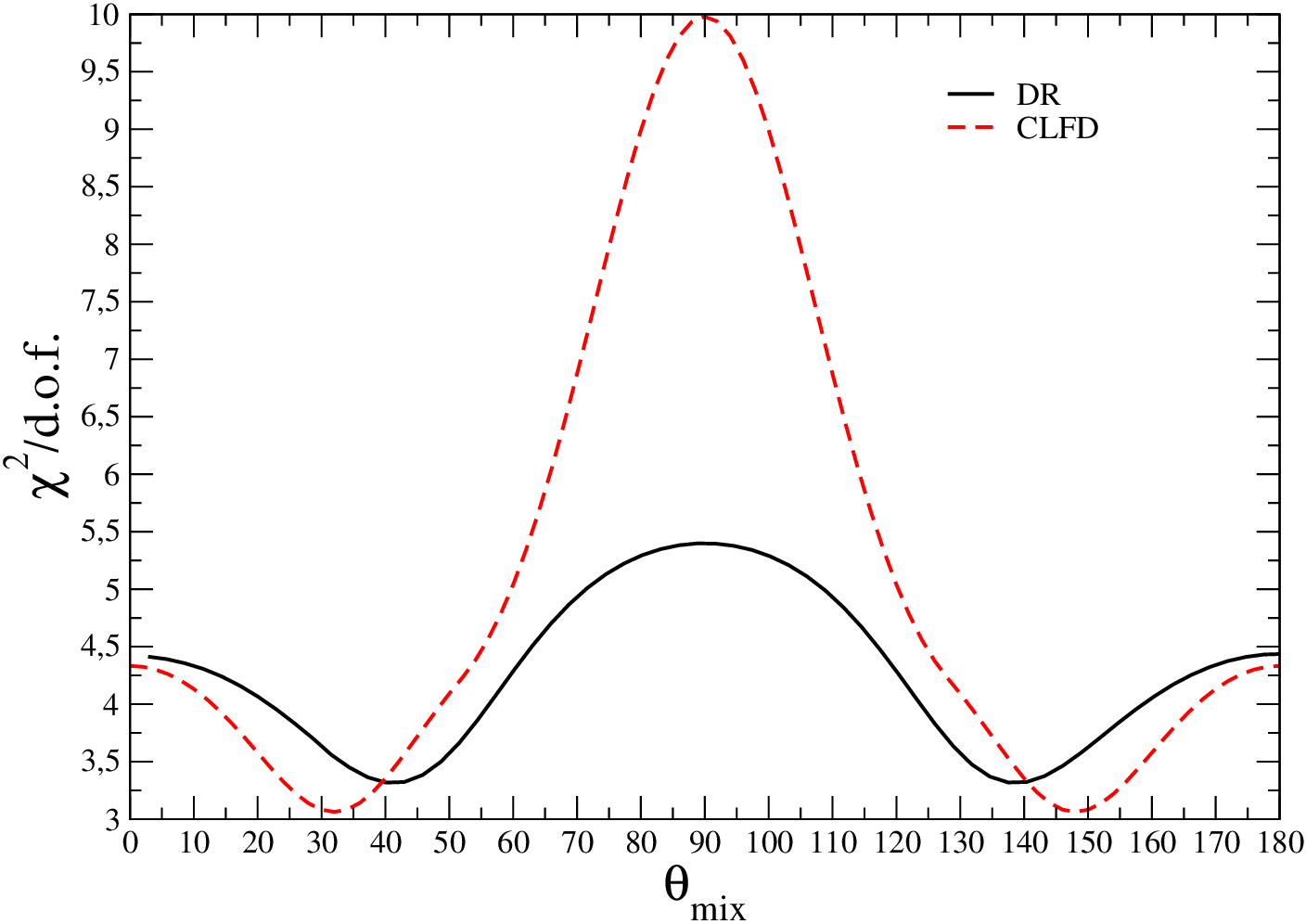}
\end{center}
\caption{The variation of $\chi^2/$d.o.f. as a function of the mixing angle $\theta_{\mathrm{mix}}$. It corresponds to the fit 1 where 12 branching ratios 
are fitted with 2 parameters. The full and dashed lines correspond to the DR and CLFD results, respectively.}
\label{spec12b2p}
\end{figure}

The stability of our fit is illustrated in Fig.~\ref{spec12b2p} for both approaches. The $\chi^2/$d.o.f. function is, in both cases  
(CLFD and DR), smooth and has well defined minima as a function of the mixing angle $\theta_{\mathrm{mix}}$. 
We find a mixing angle $\theta_{\mathrm{mix}} = 32^{\circ} \pm 4.8^{\circ}$ with the CLFD model and  
$\theta_{\mathrm{mix}} =41.3^{\circ} \pm 5.5^{\circ}$ for the DR model and the symmetric angles with respect to $90^\circ$. 
These value are in fair agreement with the ones estimated from $D_s^+\to f_0(980)\pi^+$ and $D_s^+\to \phi\pi^+$ decays~\cite{Yao:2006px}, 
which cover the rather wide range $20^\circ \lesssim \theta \lesssim 40^\circ$ and $140^\circ \lesssim \theta \lesssim 160^\circ$. 
 
The addition of an annihilation amplitude in that channel $D^0 \to f_0(980) \bar K^0$ does considerably improve the quality of the agreement with the 
complete set of data (fit 2a), as seen in Table~\ref{12b4pbr}. The results for the parameters $\nu_n$ and $\theta_{\mathrm{mix}}$ (Table~\ref{12b4ppa}) are extremely stable as compared to those of fit 1 (Table~\ref{12b2ppa}).

\begin{figure}
\begin{center}
\includegraphics*[angle=0,width=0.78\columnwidth]{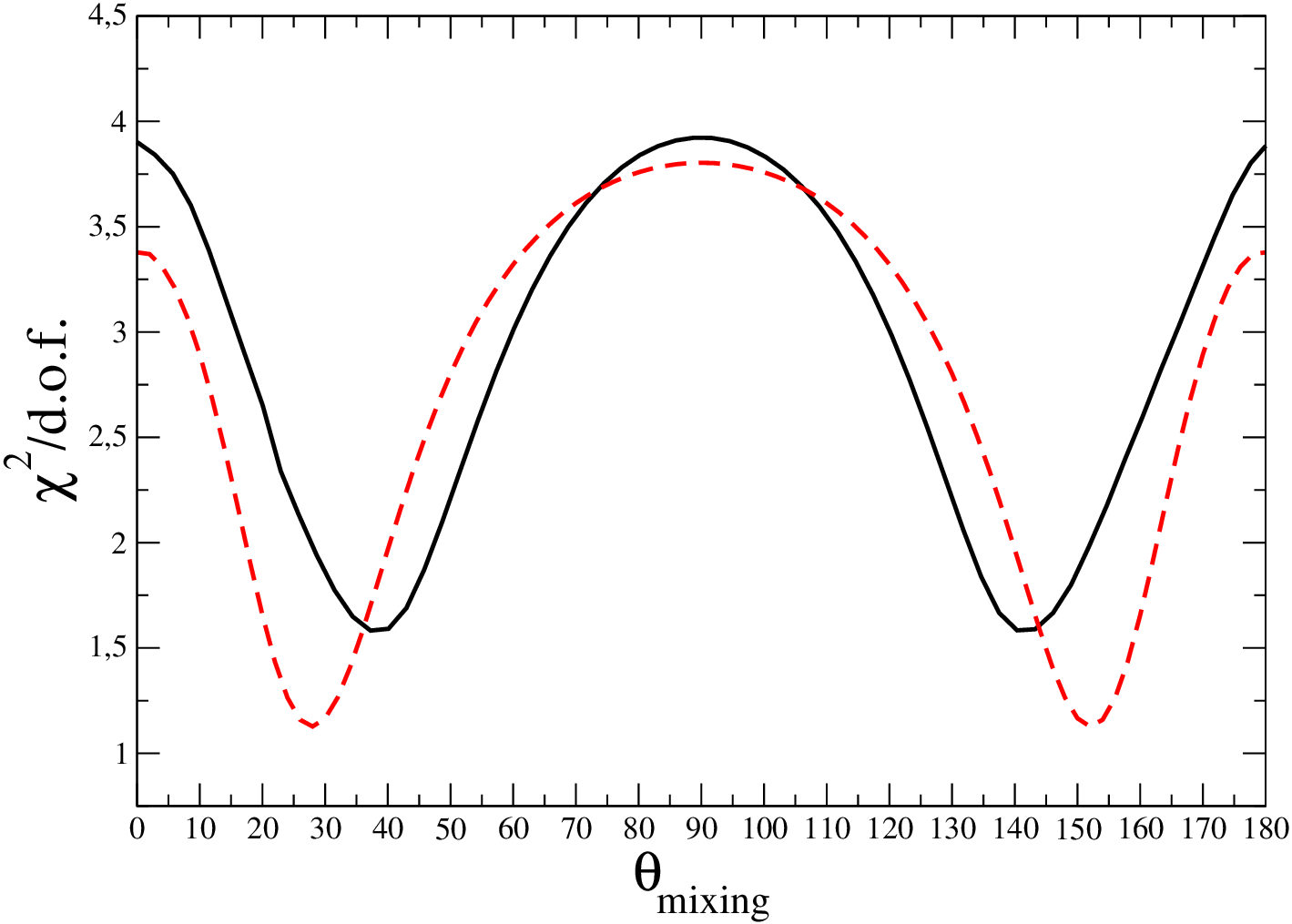}
\end{center}
\caption{Same as in Fig.~\ref{spec12b2p} but for the fit 2b where 9 branching ratios are 
fitted with 4 parameters. }
\label{spec9b4p}
\end{figure}

Finally, retaining only the nine consistent data as explained above, we obtain (fit 2b) a further improvement of the $\chi^2/$d.o.f. illustrated
in Table~\ref{9b4pbr}. The wave-function parameters (Table~\ref{9b4ppa}) for the CLFD model are stable  as compared to those in Tables~\ref{12b2ppa} 
and~\ref{12b4ppa} whereas for the DR model, 
the range parameter increases by about $20\%$ while the mixing angle remains stable. As for fit 1, the stability of fit 2b is illustrated in Fig.~\ref{spec9b4p} 
in both approaches. The $\chi^2/$d.o.f. function is, in both the CLFD and DR models, smooth and has well defined minima as a function of the mixing 
angle $\theta_{\mathrm{mix}}$. 

The prediction for $\theta_{\mathrm{mix}}$ differs by about $10^\circ$ in the two approaches. This can be explained as follows; in both approaches we employ equal Gaussian parametrizations of the vertex functions introduced in Eqs.~(\ref{defaclfd}) and~(\ref{wk}), yet the dynamics that enters the loop diagram associated with the meson normalization differs somewhat in each case. In particular, in the DR approach the condition in Eq.~(\ref{normG}) implies a vertex renormalization due to soft rescattering of the constituent quark in the vicinity of the meson pole mass $M^2$. These modifications in the calculation of the  normalization already cause differing normalization values in the case of the heavy pseudoscalars.
As seen in Table~\ref{pseudopa} in Appendix~\ref{appA}, although the values of $\nu$ are very close in both models, the normalizations are quite different. This feature of the normalization is even more apparent for the $f_0(980)$ but, in addition, the size parameters $\nu$ are an order of magnitude apart which results in different weights and ranges of the bound state vertex functions. These unequal weights enter the form factor calculations where they are compensated by the different normalizations $N_S$ in both models. However, another degree of freedom comes into play here, namely the $f_0(980)$ mixing angle whose value can also compensate the Gaussian weights and thus competes with the normalization. Since for small momentum transfers, $q^2=m_\pi^2$  and $m_K^2$, the effective form factors of CLFD and DR [see Eqs.~(\ref{effecffn}) and  (\ref{effecffs})] must be very close in order to fit the data, the product of $N_S$, the Gaussian weights and the sine or cosine of the mixing angle in the decay amplitudes  Eq.~(\ref{decayamplitudes}) must agree up to small variations inherent to a fit with two different models. The normalization $N_S$ being not equal in CLFD and DR, this results in the observed variation of about $10^\circ$ in the mixing angle.

All fits of the branching ratios  only constrain the $f_0(980)$ wave function at very small relative momenta $k^2$,  of the quark pair as given in 
Eq.~(\ref{eq:4kq2}).  Though the introduction of the annihilation amplitude  considerably improves the fit, its consequences on the scalar meson parameters are rather limited.

\subsection{$\boldsymbol{P\to S}$ transition form factors\label{sec7.1}} 

\begin{figure}
\begin{center}
\includegraphics*[width=0.77\columnwidth]{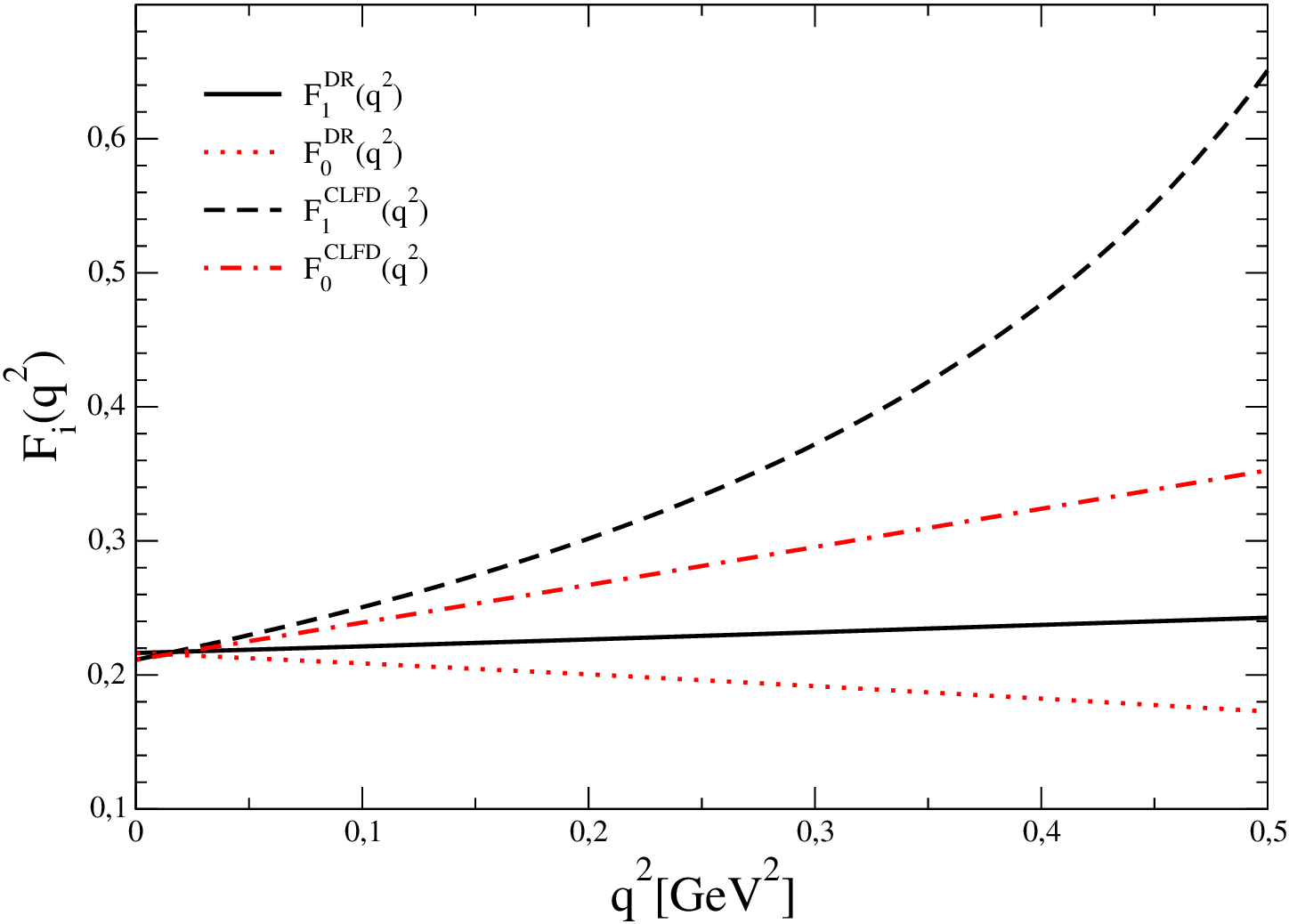}
\end{center}
\caption{Effective form factors  $F_0(q^2)$  and $F_1(q^2)$ [see Eq.~(\ref{effecffn})] calculated with the parameters of fit 2b  for $D \to f_0(980)$ transitions. In the DR model, the full and 
dotted lines correspond to $F_1(q^2)$ and $F_0(q^2)$ respectively, and similarly for the dashed and dot-dashed lines in the CLFD model.}
\label{ffDutofou}
\end{figure}

\begin{figure}
\begin{center}
\includegraphics*[width=0.77\columnwidth]{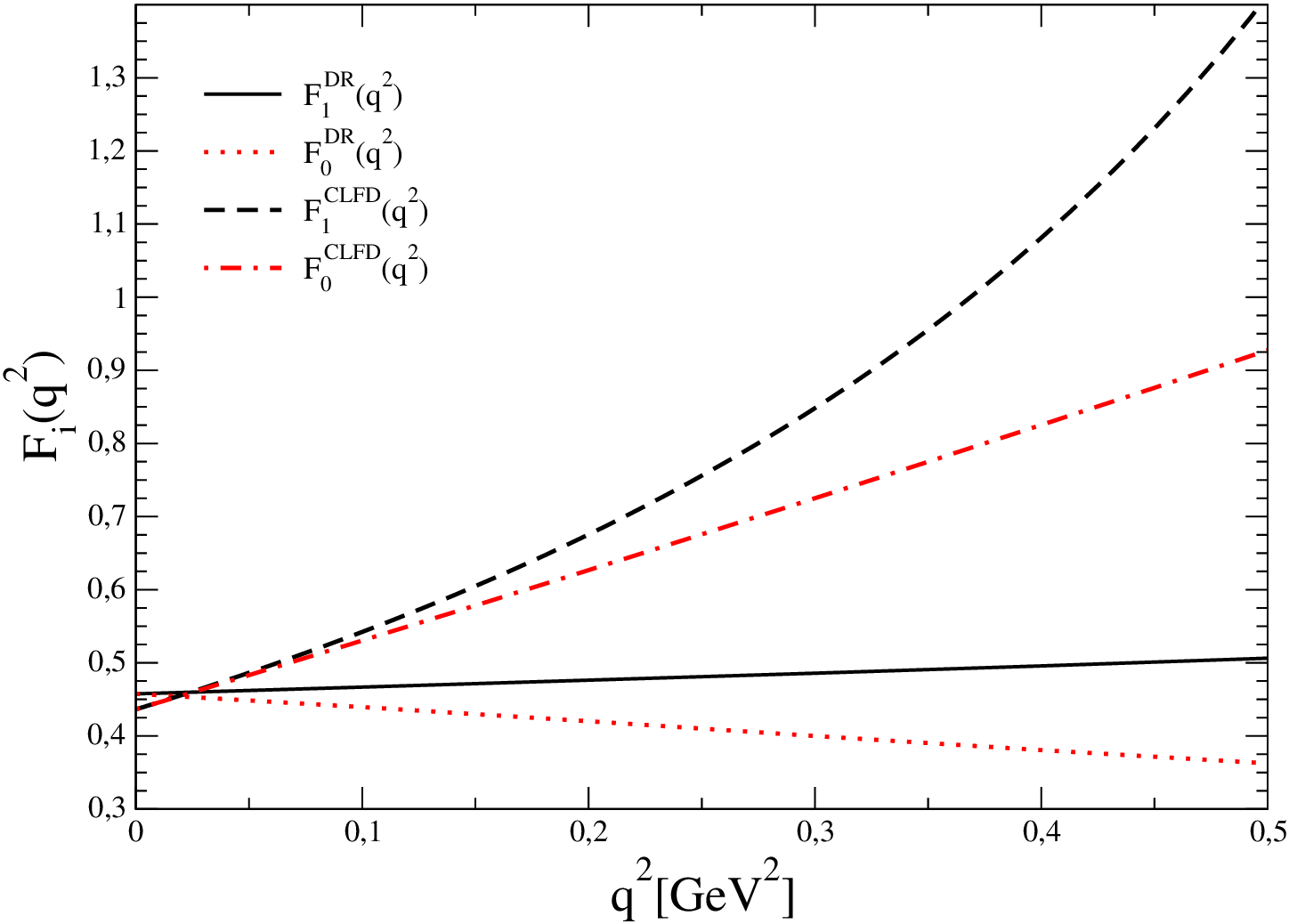}
\end{center}
\caption{Same as in Fig.~\ref{ffDutofou} but for $D_s \to f_0(980)$  transitions [see Eq.~(\ref{effecffs}) for the definition of $F_0(q^2)$ and $F_1(q^2)$]. 
\label{ffDstofos}}
\end{figure}

 With the parametrization of the scalar-meson wave function in  Table~\ref{9b4ppa}, resulting from fit 2b, we compute the pseudoscalar to scalar transition 
 form factors $D\to f_0(980)^{(n)}$, $D_s\to f_0(980)^{(s)}$ and can now predict $B\to f_0(980)^{(n)}$ and $B_s\to f_0(980)^{(s)}$. 
Indeed, with the values of Table~\ref{pseudopa} in Appendix~\ref{appA}, we can compute, employing Eqs.~(\ref{eq9.5}), (\ref{eq9.5b}), (\ref{eq9.12}) and  
(\ref{f+-full}), the $PS\to S$ transition form factors $D \to f_0(980)^{(n)}$, $D_s\to f_0(980)^{(s)}$, $B \to f_0(980)^{(n)}$  
and $B_s\to f_0(980)^{(s)}$ for any kinematically allowed momentum transfer $q^2$. In the CLFD formalism this is done for $q^2>0$  
whereas in DR these form factors are evaluated for spacelike and timelike values of $q^2$.  Since we compare the two models, we only consider
the positive range of $q^2$. The momentum-transfer dependence of the effective $F_0(q^2)$ and $F_1(q^2)$ form factors is plotted in Figs.~\ref{ffDutofou}  
and \ref{ffDstofos} for the $D$ and $D_s$ transition form factors in both models. It is worthwhile to mention  
that, in the DR formalism, the anomalous contribution to the form factors, namely the second term in Eq.~(\ref{f+-full}), only sets in for momenta of  
$q^2\gtrsim 0.6$ GeV$^2$ in $D_{(s)}\to f_0(980)$ transitions. Therefore, in the momentum range of interest here, this contribution 
is negligibly small and  only the Landau part of the integrals is of interest. To make these effective form factors readily available, we assemble in Tables~\ref{valffDtofo} and~\ref{valffBtofo} a list of their values for a few specific values of $q^2$, namely $q^2= m_\pi^2,\ m_K^2$ and $m_{\rho}^2$.
 
As can be read from Table~\ref{valffDtofo} and Fig.~\ref{ffDutofou} and~\ref{ffDstofos}, both models are in fair agreement for the range  
of timelike momenta $0<q^2\lesssim 0.1$~GeV$^2$ in the transitions $D \to f_0(980)^{(n)}$ and $D_s \to f_0(980)^{(s)}$. This is  
expected as in the fit we fix the model parameters via the effective form factor $F_0(q^2)$ for $q^2=m_\pi^2$ and $m_K^2$ barring any other changes 
in the decay amplitudes of Eq.~(\ref{decayamplitudes}). 

\begin{figure}
\begin{center}
\includegraphics*[width=0.77\columnwidth]{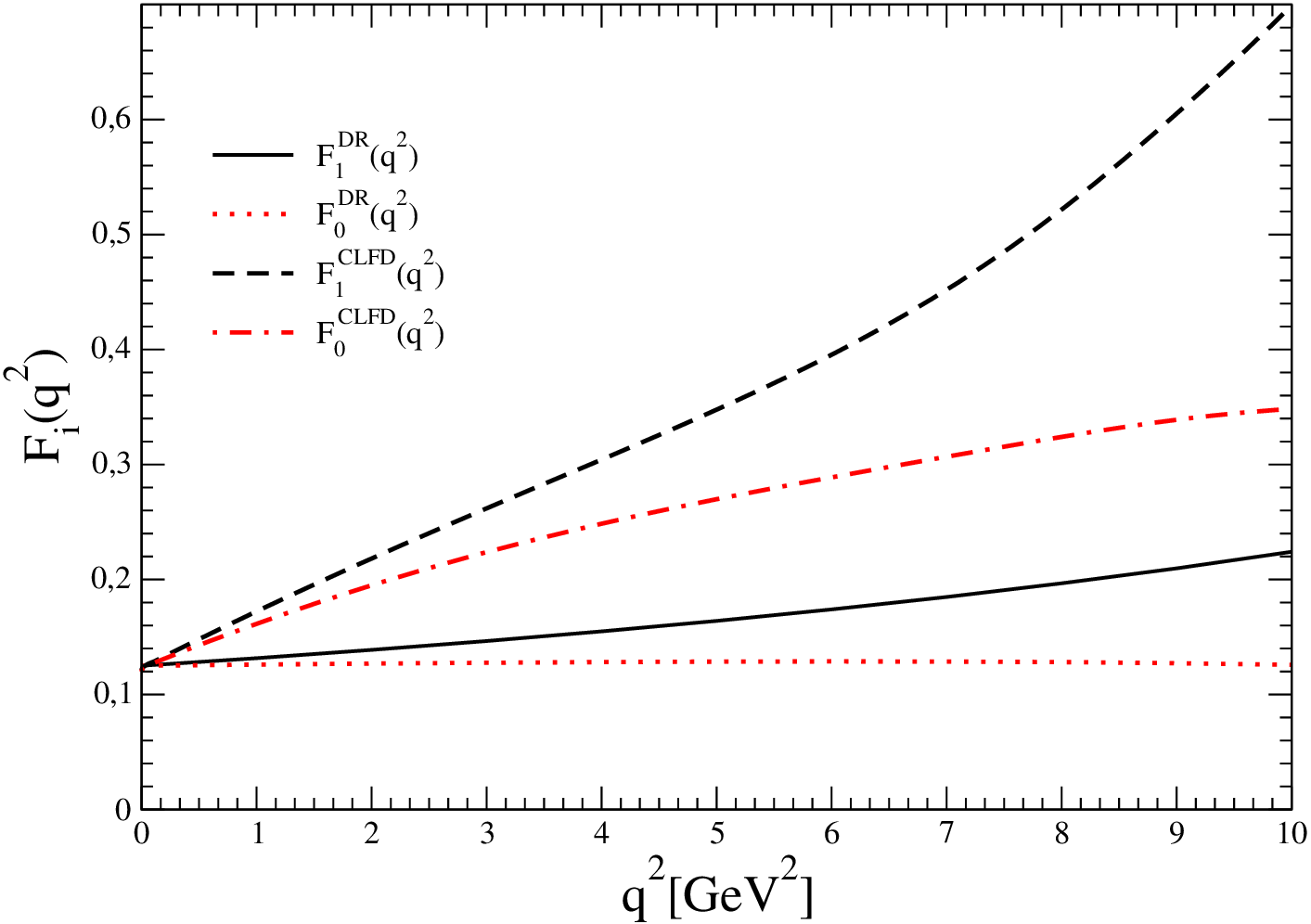}
\end{center}
\caption{Same as in Fig.~\ref{ffDutofou} but for $B \to f_0(980)$  transitions.}
\label{ffButofou}
\end{figure}

\begin{figure}
\begin{center}
\includegraphics*[width=0.77\columnwidth]{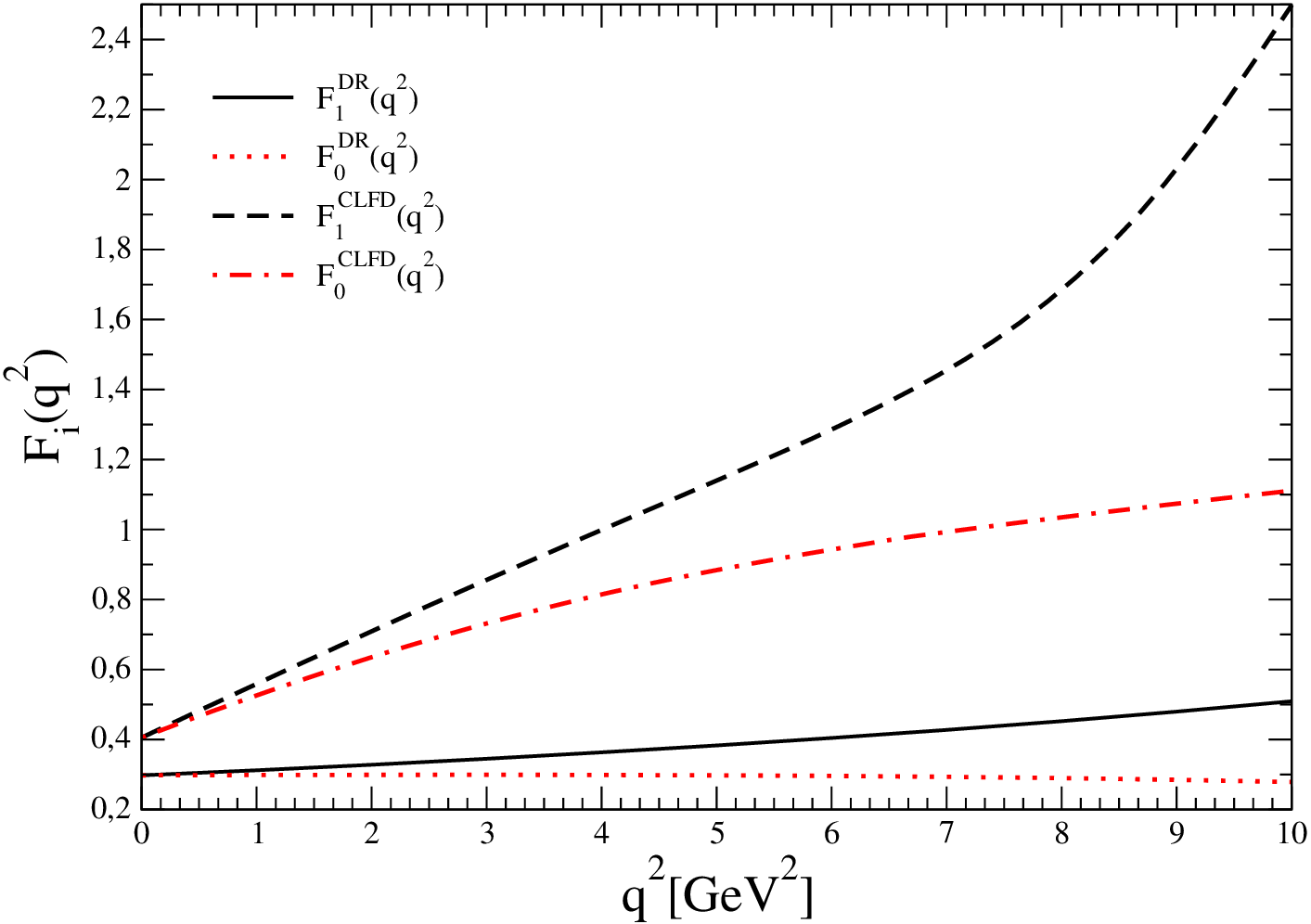}
\end{center}
\caption{Same as in Fig.~\ref{ffDstofos} but for $B_s \to f_0(980)$  transitions  [see Eq.~(\ref{effecffs})].\label{ffBstofos}}
\end{figure}

For the $B$ to scalar transitions\footnote{Our attention has been drawn by R.~Dutta~\cite{Dutta:2008xw} to a work with S. Gardner where they obtained 
similar results with the use of the constituent quark model combining heavy quark effective theory with chiral symmetry in the light quark sector.}, the kinematically allowed range is much larger than extending the momentum  transfer squared  up to $15$ GeV$^2$. Hence, once again, we do not consider contributions of the anomalous term in Eq.~(\ref{f+-full}) in the DR formalism.  The effective form factors $F_0(q^2)$ and $F_1(q^2)$ are plotted in Figs.~\ref{ffButofou} 
and~\ref{ffBstofos}. Table~\ref{valffBtofo} gives a few values at $q^2=m_{\pi}^2,\ m_{K}^2,\ m_{\rho}^2$ and $m_{D}^2$.  In Fig.~\ref{ffButofou}, for the 
$B \to f_0$ transition, one observes similar results to those obtained for the $D$ to scalar transitions, whereas for the $B_s \to f_0$, the difference 
between the DR and CLFD predictions is considerable as can be seen in Fig.~\ref{ffBstofos}.

The magnitude of the slopes for $F_0(q^2)$ and $F_1(q^2)$ point  
at different dynamical features for larger $q^2$ despite the use of similar vertex functions in both CLFD and DR. This is true in particular  
for large $q^2\simeq m_b^2$ values in $B\to f_0(980)$ transitions where one expects perturbative QCD effects to be relevant. 
It is likely that the Gaussian vertex form of the Bethe-Salpeter amplitudes which describe both the heavy pseudoscalar and the  
light(er) scalar bound states are not appropriate at large momentum transfers. In the $D$ decays, the differences are even more  
pronounced --- whereas at the maximum recoil point $q^2=0$ the DR approach values for $F_1(q^2)$ are slightly  
larger in magnitude than those from CLFD, they evolve more slowly and at $q^2_\mathrm{max}$ the CLFD predictions are considerably  
larger, as seen in Figs.~\ref{ffButofou}  and \ref{ffBstofos}. In this case, the momentum transfer range $0\leq q^2\leq 0.6$~GeV$^2$  
is lower than the meson mass $m_D^2$ and the process should be more dominated by soft physics. Therefore, the deviations  
between DR and CLFD cannot be ascribed to the behavior of the vertex functions and are intrinsic to the dynamical assumptions  
in either model. A feature of the DR model is that the function $f_-(q^2)$ decreases more rapidly than $f_+(q^2)$ increases, in  
particular for the $D\to f_0(980)$ transition form factors. This steeper slope as well as the factor $q^2/(m_D^2-m_{f_0}^2)$ which  
is larger than $q^2/(m_B^2-m_{f_0}^2)$ in Eq.~(\ref{eq9.5b}) also explain the negative slope of $F_0(q^2)$ for the  $D\to f_0(980)$  
transitions. However, the difference with the CLFD form factor prediction is striking and only in the momentum domain of the pion  
and kaon mass can agreement be found.  The problem of model dependence appears at larger momentum transfer, where various
 models yield rather different results, whereas at $q^2 =0$ Ref.~\cite{Dutta:2008xw} seems to confirm our results.
  
Regarding the general behavior of the transition form factors, in DR one observes that they are very sensitive to the function $b_{\pm}$ 
which strongly depend on the quark mass difference (Eqs.~(\ref{bpbm}) and~(\ref{B+-})). In CLFD, the form factors 
are controlled by the function $\Omega$, introduced in Eq.~(\ref{eq9.13}), which forces  $F_0(q^2)$ and $F_1(q^2)$ to behave as 
$1/(\alpha+\beta q^2)$ and therefore become very large at the kinematical limit  whenever the denominator tends to zero.

It is worthwhile to recall that quark model predictions have a constituent mass dependence causing a systematic error in the  
computation of the form factors. This is in particularly true for the light sector where it is known that the dressed-quark mass 
receives strong momentum-dependent corrections at infrared momenta, an expression of dynamical chiral symmetry breaking. 
The enhancement of the mass function in the light-quark propagators is central to the occurrence of a constituent-quark mass scale. 
On the other hand, the impact on heavy-quark propagators of chiral symmetry breaking is much less marked for $c$-quarks 
and even less so for $b$-quarks. It can be shown that  that the heavy propagator $S(p) = (\SLASH p -m_Q)^{-1}$ is justified for  
$b$-quarks and to a certain extent also for $c$-quarks \cite{Ivanov:1998ms}. Thus, in the approach of same propagators 
for light and heavy quarks, with a light constituent mass of $m_{u,d} = 0.35$~GeV, a certain mass dependent uncertainty is implicit.  
We also remind that both $D$ and $B$ mesons are lightly bound and that the bound state condition $M^2< (m_1+m_2)^2$  
is only fulfilled in the quark model if the light-quark mass is chosen to be large. However, since we make use of the features  
of confining models, this constraint does not affect our predictions. 
 
As an example, if we choose for the light-quark mass $m_{u,d}=0.25$~GeV, modifications of the form factors magnitude at larger  
$q^2$ values are not insignificant. In the DR approach, for instance, a decrease of the light-quark mass, which implies a readjustment of the  
meson parameters to fit their decay constants, $F_0(q^2)$ and $F_1(q^2)$ evolve more rapidly and overall we observe modifications 
of the order of $10\%$ for $q^2$ up to the squared kaon mass. Changes in the strange quark mass scarcely alter 
these form factors on the other hand. This observation is  
more striking for $D\to f_0(980)$ transitions, where the heavy-light quark mass differences $(m_c-m_u)^2$ and  $(m_c-m_s)^2$ are  
smaller than when a $b$-quark is involved. A proper treatment of dressed light-quark propagators should remedy this situation.

\section{Epilogue \label{sec8}} 
 
We have investigated the role of the scalar meson $f_0(980)$ in quasi-two-body decays of $D_{(s)}$ and $B_{(s)}$ mesons focussing  
on the weak transition form factors $D_{(s)}\to f_0(980)^{(s)}$ and $B_{(s)}\to f_0(980)^{(s)}$, which are of particular interest to flavor physics.  
In order to obtain a consistent parametrization of the $f_0(980)$ wave function, we first applied a simple factorization ansatz to these  
$D$ decays where the approach is reasonable.   Here, the quasi-two body $D_{(s)}\to f_0(980) P$ branching ratios are deduced from the 
experimental ones for  $D_{(s)} \to \pi\pi\pi,\ \bar KK\pi$ and the knowledge of the $f_0(980)\to \pi^+\pi^-$ and $f_0(980)\to K^+K^-$ branching fractions.  
Once the scalar-meson parameters are determined by fitting the  matrix element $D_{(s)}\to f_0(980)^{(s)}$ to experimental data, they are readily 
available for other flavor changing matrix elements  involving the $b$-quark although in that case the approach is on less firm grounds.
The short-distance physics in the factorization is known from perturbation theory applied to the operator product expansion and codified 
in terms of Wilson coefficients. The long-distance effects concern two sets of form factors; namely the experimentally known decay constants 
and the heavy-to-light transition form factors. The latter are nontrivial objects which involve quark as well as hadron degrees of freedom.  
In our approach, we have modeled these form factors with triangle diagrams (at the tree level) in the impulse approximation. The mesonic Bethe-Salpeter  
amplitudes are described by Gaussian two-quark vertex functions which introduce size parameters. In the case of the scalar meson, we  
also need a mixing angle between the strange and nonstrange components of its wave function for which we assume the simplest possible  
quark structure. That is to say, we neglect higher Fock states or possible hadronic dressings which may enrich the $\bar qq$ state with other  
components such as $| \bar KK\rangle$, $|  \pi \eta \rangle$ {\em etc\/}. in order to perform an actual calculation.  
As noted previously in Secs.~I and II, a consequence of the mixing is the presence of a strange component in the $\sigma$ or $f_0(600)$ state, strange content which does not seem to be experimentally observed.  
A specific discussion of the structure of this broad state is outside the scope of the present study and  would require, as we just pointed out, to  work beyond the simplest two-quark structure.
 
In this work, we have examined two different but explicitly covariant approaches to establish the model dependance of the form factors.  
In both model calculations, the impulse approximation is used and quark masses as well as dynamical assumptions are the identical, though  
certain kinematical aspects differ. In particular, in the DR approach internal quarks are put on-mass shell and the amplitudes are expressed  
as double dispersive integrals of the triangle diagram's discontinuity over initial and final mass variables. In contrast to the DR approach, in the CLFD  
calculation the integration is performed over the internal loop momenta. Moreover, even though the Bethe-Salpeter amplitudes of the  
$D_{(s)}$ and $f_0(980)$ have identical Gaussian forms, the meson vertex normalization is not identical in both models.  
 
These differences may be the origin for certain discrepancies we find in our  results. In fitting the set of experimental $D_{(s)}\to f_0(980) P$  
branching ratios, we do obtain  similar values for the mixing angle.
 Overall, the fit quality  
is comparable and rather good given the large experimental errors. However, while  at small momentum transfer, around the light meson  
masses $m_\pi^2$ and $m_K^2$, we find very similar transition form factors, for  larger values of $q^2$ where no experimental constraints  
exist the discrepancy is obvious. In the case of $B_{(s)}\to f_0(980)$ transitions,  stronger deviations between both models are observed. For  
the $D_{(s)}\to f_0(980)$ transitions, the discrepancy  is already obvious for $q^2\lesssim m_K^2$ as seen, in particular, in the different  
slopes of $F_0(q^2)$ obtained in DR and CLFD.  This is also a hint that the constituent quark model may be reliable solely for a  
certain domain of $q^2$.  

Clearly, dynamical aspects of QCD, such as running quark masses, are important in  the computation of these form factors.  
 
 Furthermore, the parametrization of the heavy mesons depends on the precise knowledge of the pseudoscalar decay  
constant. As confinement is only approximately achieved and dynamical chiral symmetry breaking not realized in either model calculation,  
some of the uncertainty defies any quantification. When these formalisms are applied to calculations which can be compared  
to observables such as decay constants, typical deviations from the experimental values are of the order of $10\%-15\%$. 
Given the large errors in the experimental $D_{(s)}\to \pi\pi\pi,\ \bar KK \pi$ branching fractions and the still elusive structure of 
the scalar $f_0(980)$, assuming a $\bar qq$ composition, this provides a lower bound of our theoretical  
error\footnote{which can be roughly tested by varying the size parameters and mixing angle within the error  
ranges shown in Table~\ref{9b4ppa}} which we estimate to be of the order of $25\%$. 
 
Nonetheless, we consider that there is a domain of validity for these models which overlaps with the typical momentum transfers  
$q^2$ that occur in leptonic as well as nonleptonic weak decays of  $D_{(s)}$ and $B_{(s)}$ mesons. The present study provides a  
first calculation of heavy pseudoscalar to scalar meson transition form factors at the exact momentum transfer values  
$q^2=m_\pi^2,\ m_K^2,\ m_\rho^2$ and $m_D^2$ without resorting to any extrapolation. Surely, this work leaves plenty of room 
for improvement; obviously a better understanding of the scalar-meson structure is of foremost concern, but a more genuine 
realization of confinement and dynamical chiral symmetry breaking is also desirable.

\subsection*{Acknowledgments}
We  warmly thank D.~Melikhov for fruitful exchanges during the course of this work. B.~E. acknowledges 
financial support from {\em Marie Curie International Reintegration Grant\/} N$^\circ$~516228 and 
valuable communication with C.~D.~Roberts, J.~P.~B.~C~de Melo and T.~Frederico. This work was partly supported by an {\em IN2P3-CNRS\/} 
theory grant for the project ``{\em Contraintes sur les phases fortes dans les d\'esint\'egrations hadroniques des m\'esons B\/}" and
by the Department of Energy, Office of Nuclear Physics, Contract  No.~DE-AC02-06CH11357.

\newpage
\renewcommand{\theequation}{A\arabic{equation}}
\setcounter{equation}{0}

\appendix

\section{Pseudoscalar mesons in the quark model\label{appA}}

\subsection{CLFD}

For a pseudoscalar particle composed of an antiquark  and a quark, of mass $m_1$ and $m_2$ respectively, the general structure of the two-body bound 
state has the form:
\begin{equation}\label{eq8.1}
\psi_P^{(q q^{'})}=N_P \;  \phi_P^{(q q^{'})}, \;\; {\rm with} \;\; \phi_{P}^{(q q^{'})}=\frac{1}{\sqrt{2}}\bar{u}(k_2)A^{(q q^{'})}_P(x,{\bf R}_{\perp}^2)
\gamma_5\  v(k_1),
\end{equation}
where  $v(k_{1})$ and  ${\Bar u}(k_{2})$  are the usual Dirac spinors, and $A^{(q q^{'})}_P(x,{\bf R}_{\perp}^2)$ is the  scalar component of the wave function written as 
\begin{equation}\label{pseudoeq}
A^{(q q^{'})}_P(x,{\bf R}_{\perp}^2)=A^{(q q^{'})}_P({\bf k}^2)=\exp(-4 \nu {\bf k}^{2}/m_{12}^2),
\end{equation}
  where $N_P$ and $\nu$ are parameters to be determined 
by comparison with experimental 
data; the  reduced mass is $m_{12}= m_1m_2/(m_1+m_2)$ and  ${\bf k}^2$ is given by Eq.~(\ref{grask2}). For the pseudoscalar mesons we 
make use of the experimentally well established values for their decay constant.

In CLFD the normalization condition  for a pseudo scalar meson of zero total angular momentum reads as follows :
\begin{eqnarray}
1 = \int_{(x,\tilde\theta,{\bf R}_{\perp})} D(x,\tilde\theta,{\bf R}_{\perp}) 
\sum_{\lambda_{1} \lambda_{2}} \psi_{\lambda_{1} \lambda_{2}}^{(qq^{'})} 
\psi_{\lambda_{1} \lambda_{2}}^{(qq^{'})\star},
\end{eqnarray}
where, in close analogy with Eqs.~(\ref{sumlambda12}), one has,
\begin{align}
\sum_{\lambda_1,\lambda_2} \psi_{\lambda_1,\lambda_2}^{(qq^{'})}\  \psi_{\lambda_1,\lambda_2}^{(qq')\dagger}  
 & =  \dfrac{N_P^2}{2}\mathrm{Tr}
  \left[
 (\SLASH k_2+m_2) A^{(q q^{'})}_P(x,{\bf R}_{\perp}^2) \gamma_5 (\SLASH k_1-m_1) A^{(q q^{'})}_P(x,{\bf R}_{\perp}^2) \gamma_5
 \right],
\end{align}
so that, finally,
\begin{eqnarray}\label{eqnormps}
1 = N_P^2 
\int_{(x,\tilde\theta,{\bf R}_{\perp})} D(x,\tilde\theta,{\bf R}_{\perp})
 \Biggl\{ \left[ \frac{{\bf R}_{\perp}^{2}+(x m_{2}+
(1-x) m_{1})^{2}}{x(1-x)} \right] \bigl[A^{(q q^{'})}_P(x,{\bf R}_{\perp}^2)\bigr]^2 \Biggr\},
\end{eqnarray}
where one recalls that $D(x,\tilde\theta,{\bf R}_{\perp})$ is the invariant phase
 space element already defined in Eq.~(\ref{eq9.19}).

\subsection{Dispersion approach}

Similarly, the the two-body bound state for pseudoscalar meson is given here by
\begin{equation}
\label{vertexps}
\langle P(k_1,k_2) | \bar QQ\rangle = \frac{\bar Q^{a}(-k_2)i\gamma_5 Q^{a}(k_1)}{\sqrt{N_C}}\ G_v(s),
\end{equation}
where $Q^{a}(k_1,m_1)$ represents the spinor state of the constituent quark of
 color $a$ and $N_C=3$ the number of quark colors.
Since for a confining potential the strong interaction does not produce a pole
 at $s=M^2$ in the physical region (in the harmonic
oscillator approximation of the quark model the Gaussian functions are smooth),
 the vertex function $G_v(s)$ can be related, as in Eq.~(\ref{phiG}), to
a wave function representation of the form
\begin{equation}
\psi_P(s)=G_v(s)/(s-M^2)= N_P \phi_P(s),
\end{equation}
where $N_P$ is a normalization factor and 
\begin{equation}\label{psidr}
\phi_{P}(s)=\frac{\pi}{\sqrt{2}}\ \frac{\sqrt{s^2-(m_2^2-m_1^2)^2}}{\sqrt{s-(m_2-m_1)^2}}
\ \frac{1}{s^{3/4}}\ w(k). 
\end{equation}
In Eq.~(\ref{psidr}), the function $w(k)$ is chosen to be
\begin{equation}
w(k)= \exp\left(-4\nu k^2/m_{12}^2\right),
\end{equation}
where $m_{12}$ is again the reduced mass. As in CLFD, we determine 
the normalization, $N_P$, and fit the size parameter $\nu$ so as to reproduce the experimental decay constants. In the dispersion 
approach the relativistic normalization Eq.~(\ref{normG}), by 
the appropriate choice for the wave function, reduces to the simple integral
\begin{equation}\label{eqnormps2}
 1= N_P^2 \int_0^\infty w^2(k)k^2 dk, \;\; {\rm with}\;\; N_P=\frac{2}{\pi^{1/4}}\Bigl(\frac{8 \nu}{m_{12}^2} \Bigr)^{3/4}.
\end{equation}

\subsection{Decay constant of the pseudoscalar mesons}
 According to the usual definition, the decay amplitude  is  ${\Xi}^{\mu}= \langle0 |
 J^{5\mu} | P \rangle$ where $J^{5\mu}$ is the axial current. 
 Since our formulation is explicitly covariant, we can decompose ${\Xi}^{\mu}$
 in terms of all momenta available in our system, i.e. the incoming meson momentum
$p^{\mu}$ and $ \omega^{\mu}$. 
We have therefore:
\begin{equation}\label{eq8.20}
{\Xi}^{\mu}= f_{P}\ p^{\mu} + \mathcal{B}\  \omega^{\mu},
\end{equation}
where  $f_{P}$ is the physical decay constant. In an exact calculation of
 $\Xi_{\mu}$, $\mathcal{B}$ should be  zero.
Since $\omega^2=0$, the decay constant  can  easily be obtained according to:
\begin{equation}\label{eq8.21}
 f_{P} = \frac{\Xi \sdot  \omega}{\omega \sdot p}.
\end{equation}
Using the diagrammatic rules of CLFD, we can calculate $\Xi^\mu$ and including
 color factors, one gets,
\begin{equation}\label{eq8.22}
{\Xi}^{\mu}=\sqrt{3} N_P \int_{(x,\tilde\theta,{\bf R}_{\perp})} 
D(x,\tilde\theta,{\bf R}_{\perp})
{\rm Tr} \left[ -\overline{ \gamma^{\mu}\gamma^5}(\SLASH{k_{2}} 
+m_{2}) \frac{1}{\sqrt{2}}A^{(q q^{'})}(x,{\bf R}_{\perp}^2) \gamma^5(m_{1}
- \SLASH{k_{1}}) \right],
\end{equation}
 where the notation $\overline O$ is defined as usual by
$ \overline{O} = \gamma^{0} O^{\dagger} \gamma^{0}$. The decay constant is therefore given by:
\begin{equation}\label{fpclfd}
f_P = 2 \sqrt{6}N_P \int_{(x,\tilde\theta,{\bf R}_{\perp})} D(x,\tilde\theta,{\bf R}_{\perp})
\Bigl[m_{1}(1-x)+m_{2}x \Bigr]A^{(q q^{'})}(x,{\bf R}_{\perp}^2).
\end{equation}

Similarly, in the dispersion approach, taking into account soft rescatterings 
of constituent quarks, one obtains
a series of dispersion graphs that involve the spectral density $\rho_P(s,m_1,m_2)$
 of the Feynman quark antiquark loop graph given in Eq.~(\ref{rhops}).
These graphs yield the following expression for the pseudoscalar decay constant~\cite{Melikhov:2001zv}
\begin{eqnarray}\label{fpdr}
 f_P & = &  N_P\ \sqrt{N_C} \int^\infty_{(m_1+m_2)^2} \,  \frac{ds}{\pi} \, \frac{m_1+m_2}{s} \,
\rho_P(s,m_1,m_2)\ \phi_P(s).
\end{eqnarray}

Then applying the normalization condition with the  decay constants as constraints for 
modelling, we obtain the parameters listed in Table~\ref{pseudopa}.

\renewcommand{\theequation}{B\arabic{equation}}
\setcounter{equation}{0}  
\section{Details of spectral densities}\label{appB}
Note that the double spectral densities for the pseudoscalar to scalar transition form factors 
$\Delta_\pm(s_1,s_2,q^2;m_1,m_2,m_3)$ in Eqs.~(\ref{fplusmoins}) and 
(\ref{doubles}) may be obtained from Melikhov~\cite{Melikhov:2001zv} 
(section IIC) by the substitution $m_1$ into $-m_1$. This substitution is 
the consequence of the different expressions of the operators in 
the trace entering in Eq.~(\ref{doubles}) in the case of pseudoscalar to
 scalar transition and in the case of pseudoscalar to pseudoscalar 
transition~\cite{Melikhov:2001zv}. Nevertheless, for completeness, we give
 here the explicit expression. One has 
\begin{equation}
\label{deltaplusmoins}
\Delta_\pm(s_1,s_2,q^2;m_1,m_2,m_3)=
\frac{B_\pm(s_1,s_2,q^2)}{\lambda(s_1,s_2,q^2)}\ 
\Delta(s_1,s_2,q^2;m_1,m_2,m_3),
\end{equation}
where
\begin{eqnarray}\label{Bpm}
  B_+(s_1,s_2,q^2) & = & b_+(s_1,s_2,q^2) \left [ a(s_1,m_2,m_3)+a(s_2,m_3,-m_1)
-a(q^2,-m_1,m_2) \right] \nonumber \hspace*{1.5cm} \\
                   & + & a(q^2,-m_1,m_2)\lambda(s_1,s_2,q^2),  \\
  B_-(s_1,s_2,q^2) & = & b_-(s_1,s_2,q^2) \left [a(s_1,m_2,m_3)+a(s_2,m_3,-m_1)
-a(q^2,-m_1,m_2) \right] \nonumber \\
                   & + & \left[a(s_2,m_3,-m_1)-a(s_1,m_2,m_3)
\right]\lambda(s_1,s_2,q^2), 
\end{eqnarray}
with\footnote{In the expression of $b_-(s_1,s_2,q^2)$ given by Melikhov~\cite{Melikhov:2001zv}  (see his Eq.~(2.76)) there is a misprint: the relative sign between 
the two term should be $+$ as here in Eq.~(\ref{B+-}).}
\begin{eqnarray}\label{bpbm}
  b_+(s_1,s_2,q^2) & = & -q^2(s_1+s_2-q^2+m_1^2+m_2^2-2m_3^2) -(m_1^2-m_2^2)(s_1-s_2),  \\
  b_-(s_1,s_2,q^2) & = & (m_1^2-m_2^2)(2s_1+2s_2-q^2)  \nonumber \\
                            & & +\,  (s_1-s_2) (s_1+s_2-q^2+m_1^2+m_2^2-2m_3^2), \label{B+-}  
\end{eqnarray}
with $a(x,y,z)=x-(y-z)^2$. Furthermore,
\begin{equation}
\label{delta}
\Delta(s_1,s_2,q^2;m_1,m_2,m_3)=
\frac{\theta\left(b_+^2(s_1,s_2,q^2)-\lambda(s_1,s_2,q^2)\lambda(q^2,m_1^2,m_2^2)\right)}
{16\lambda^{1/2}(s_1,s_2,q^2)}.
\end{equation}
The allowed intervals for the integration variables $s_1$ and $s_2$ are obtained by solving the step $\theta$-function of Eq.~(\ref{delta}),
\begin{eqnarray}
\label{sinterval}
s_2 & > & (m_1+m_3)^2, \\
s_1^-(s_2,q^2) & < & s_1 \,\, < \,\,  s_1^+(s_2,q^2),
\end{eqnarray}
with
\begin{multline}
\label{s+-}
s_1^\pm(s_2,q^2)=\frac{s_2\ (m_1^2+m_2^2-q^2)+q^2(m_1^2+m_3^2)-(m_1^2-m_2^2)(m_1^2-m_3^2)}
{2m_1^2}\\
\pm\ \frac{\lambda^{1/2}(s_2,m_3^2,m_1^2)\lambda^{1/2}(q^2,m_1^2,m_2^2)}
{2m_1^2}.
\end{multline}
The solution of the equation
\begin{eqnarray}
s_1^R=(\sqrt{s_2}+\sqrt{q^2})^2=s_1^-(s_2,q^2),
\end{eqnarray}
which reduces to
\begin{eqnarray}
s_2 + \frac{q^2+m_1^2-m_2^2}{\sqrt{q^2}} \sqrt{s_2}+m_1^2-m_3^2=0,
\end{eqnarray}
so therefore the limit $s_2^0(q^2)$ appearing in Eq.~(\ref{f+-full}) is
\begin{equation}
\label{s20}
\sqrt{s_2^0(q^2)}=-\ \frac{q^2+m_1^2-m_2^2}{2\sqrt{q^2}}+\sqrt{\left ( \frac{q^2+m_1^2-m_2^2}{2\sqrt{q^2}} \right )^{\!\!2}+\! m_3^2-m_1^2 }.
\end{equation}
Note that  in Eqs.~(\ref{deltaplusmoins}) to (\ref{B+-}) we have introduced, following Melikhov, a lightened writing for the functions $B_{\pm}(s_1,s_2,q^2)$ 
and $b_{\pm}(s_1,s_2,q^2)$ which, we stress, depend parametrically on the quark masses $m_1, \  m_2$ and  $m_3$. This is obviously the case also for $s_1^\pm(s_2,q^2)$ and 
$s_2^0(q^2)$.



\newpage

%
%
\setcounter{table}{0}
\begin{table}
\begin{center}
\begin{tabular}{|c|c|c|c|c|c|}  
\hline
  Channel &  $\mathcal{BR}$ Exp. &\;\; $\mathcal{BR}$ Th. (DR)\;\; &\;\; $\chi^2$ \;\;&\;\; $\mathcal{BR}$ Th. (CLFD)\;\; &\;\; $\chi^2$\;\;\\
\hline\hline
\;\;  $D^+ \to f_0 \pi^+$ \;\;&&&&&\\
  (E791 [22]) & \;\;$(3.80\pm 1.17)\times 10^{-4}$  \;\;& $2.7 \times 10^{-4}$ & 0.88 & $2.90 \times 10^{-4}$& 0.58\\
\hline 
\;\;  $D^0 \to f_0 \bar{K}^0$ \;\;&&&&&\\
  (ARGUS [19]) & \;\;$(6.40\pm 2.07)\times 10^{-3}$\;\; &  $8.32 \times 10^{-5}$ & 9.30 &$1.86 \times 10^{-4}$ &9.00\\
  (CLEO [17]) & \;\;$(5.00\pm 1.52)\times 10^{-3}$ \;\; &  $8.32 \times 10^{-5}$ & 10.37 & $1.86 \times 10^{-4}$&9.94\\
  (BABAR [20]) & \;\;$(9.60\pm 8.55)\times 10^{-3}$ \;\; &  $8.32 \times 10^{-5}$ & 1.24 & $1.86 \times 10^{-4}$&1.21\\
\hline
\;\;  $D^+ \to f_0 K^+$\;\; &&&&&\\
  (FOCUS [18])& \;\;$(3.07\pm 1.65)\times 10^{-4}$ \;\;   & $1.43 \times 10^{-5}$ & 3.16 & $3.26 \times 10^{-5}$&2.77\\
  (FOCUS [18])& \;\;$(1.22\pm 0.75)\times 10^{-4}$\;\; & $1.43 \times 10^{-5}$ &  2.04 & $3.26 \times 10^{-5}$&1.41\\
\hline
\;\;  $D_s^+ \to f_0 \pi^+$\;\; &&&&&\\
  (E687 [24,25])& \;\;$(3.92\pm 2.63)\times 10^{-2}$ \;\; & $1.43 \times 10^{-2}$ & 0.89  & $1.42 \times 10^{-2}$&0.89\\
  (E791 [21])& \;\;$(1.14\pm 0.38)\times 10^{-2}$ \;\;& $1.43\times 10^{-2}$ & 0.56  & $1.42 \times 10^{-2}$&0.56\\ 
  (FOCUS [18])& \;\;$(1.90\pm 0.61)\times 10^{-2}$\;\; & $1.43 \times 10^{-2}$ & 0.58  & $1.42 \times 10^{-2}$&0.57\\
  (FOCUS [18,29])& \;\;$(5.60\pm 3.08)\times 10^{-2}$ \;\; & $1.43 \times 10^{-2}$ & 1.82  & $1.42 \times 10^{-2}$&1.82\\
\hline
\;\;  $D_s^+ \to f_0 K^+$ \;\;&&&&&\\
  (FOCUS [18])&\;\; $(2.24\pm 1.49)\times 10^{-3}$\;\;  & $0.77 \times 10^{-3}$ & 0.96 & $2.13 \times 10^{-3}$&0.01\\
\hline
\;\;  $D^0 \to f_0 \pi^0$\;\; &&&&&\\
  (CLEO [46,47])& \;\;$(1.10\pm 0.97)\times 10^{-6}$ \;\; & $2.20 \times 10^{-6}$ & 1.31 & $2.41 \times 10^{-6}$&1.84\\
\hline
\end{tabular}
\end{center}
\caption{Fit 1 (12 branching ratios and 2 parameters): comparison of experimental with theoretical branching ratios. 
The fit parameters are found in Table~\ref{12b2ppa} and a best fit yields $\chi^2/\mathrm{d.o.f.}=33.25/(12-2)=3.33$ with DR 
and $\chi^2/\mathrm{d.o.f.}=30.63/(12-2)=3.06$ with CLFD.}
\label{12b2pbr}
\end{table}
\begin{table}
\begin{center}
\begin{tabular}{|c|cc||cc|}  
\hline
  & $\nu_n$ &  $\theta_{\mathrm{mix}}$ &    $\nu_s$ &  $N_S$ \\
\hline \hline
\;\; CLFD \;\;  & \;\;  $(3.20 \pm 0.40)\times 10^{-3}$ \;\;&\;\; $ 32.0^\circ \pm 4.8^\circ$ \;\;&\;\; $(4.64 \pm 0.58)\times 10^{-3}$\;\; &\;\; 2.00 \;\;   \\
\hline \hline
  DR      & $0.014\pm 0.012$   & $ 41.3^\circ \pm 5.5^\circ$ & $0.021\pm 0.017$  \;\;& 3.41  \\
\hline
\end{tabular}
\end{center}
\caption{The scalar-meson parameters, $\nu_n$ and $\theta_{\mathrm{mix}}$, obtained in the CLFD and  DR approaches with fit 1 (see Table~\ref{12b2pbr}). 
Note 
that $\nu_s$ and $N_S$ are  given by  Eq.~(\ref{sizecorr}) and by Eqs.~(\ref{eq8.19}) or~(\ref{dispnorm}),  respectively.}
\label{12b2ppa}
\end{table}

\begin{table}
\begin{center}
\begin{tabular}{|c|c|c|c|c|c|}  
\hline
  Channel &  $\mathcal{BR}$ Exp. &\;\; $\mathcal{BR}$ Th. (DR)\;\; &\;\; $\chi^2$ \;\;&\;\; $\mathcal{BR}$ Th. (CLFD) \;\;&\;\; $\chi^2$\;\;\\
\hline\hline
\;\;  $D^+ \to f_0 \pi^+$\;\; &&&&&\\
  (E791 [22]) &\;\; $(3.80\pm 1.17)\times 10^{-4}$\;\;  & $2.64 \times 10^{-4}$ & 0.97 & $2.73 \times 10^{-4}$& 0.80\\
\hline 
\;\;  $D^0 \to f_0 \bar{K}^0$\;\; &&&&&\\
  (ARGUS [19]) & \;\;$(6.40\pm 2.07)\times 10^{-3}$\;\; &  $5.58 \times 10^{-3}$ & 0.16&$5.57 \times 10^{-3}$ &0.15\\
  (CLEO [17]) & \;\;$(5.00\pm 1.52)\times 10^{-3}$ \;\; &  $5.58 \times 10^{-3}$ & 0.15& $5.57 \times 10^{-3}$&0.14\\
  (BABAR [20])& \;\;$(9.60\pm 8.55)\times 10^{-3}$ \;\; &  $5.58 \times 10^{-3}$ & 0.22 & $5.57 \times 10^{-3}$&0.22\\
\hline
\;\;  $D^+ \to f_0 K^+$\;\; &&&&&\\
  (FOCUS [18])&\;\; $(3.07\pm 1.65)\times 10^{-4}$ \;\;   & $1.40 \times 10^{-5}$ & 3.16 & $3.09 \times 10^{-5}$&2.81\\
  (FOCUS [18])&\;\; $(1.22\pm 0.75)\times 10^{-4}$\;\; & $1.40 \times 10^{-5}$ &  2.05 & $3.09 \times 10^{-5}$&1.46\\
\hline
\;\;  $D_s^+ \to f_0 \pi^+$\;\; &&&&&\\
  (E687 [24,25])& \;\;$(3.92\pm 2.63)\times 10^{-2}$ \;\; & $1.44 \times 10^{-2}$ & 0.88  & $1.43 \times 10^{-2}$&0.89\\
  (E791 [21])& \;\;$(1.14\pm 0.38)\times 10^{-2}$\;\; & $1.44\times 10^{-2}$ & 0.61  & $1.43 \times 10^{-2}$&0.58\\ 
  (FOCUS [18])& \;\;$(1.90\pm 0.61)\times 10^{-2}$\;\; & $1.44 \times 10^{-2}$ & 0.54  & $1.43 \times 10^{-2}$&0.56\\
  (FOCUS [18,29])& \;\;$(5.60\pm 3.08)\times 10^{-2}$\;\;  & $1.44 \times 10^{-2}$ & 1.81  & $1.43 \times 10^{-2}$&1.81\\
\hline
\;\;  $D_s^+ \to f_0 K^+$\;\; &&&&&\\
  (FOCUS [18])& \;\;$(2.24\pm 1.49)\times 10^{-3}$ \;\; & $7.78 \times 10^{-4}$ & 0.95 & $2.14 \times 10^{-3}$&0.01\\
\hline
\;\;  $D^0 \to f_0 \pi^0$\;\; &&&&&\\
  (CLEO [46,47])& \;\;$(1.10\pm 0.97)\times 10^{-6}$ \;\; & $2.16 \times 10^{-6}$ & 1.20 & $2.28 \times 10^{-6}$&1.49\\
\hline
\end{tabular}
\end{center}
\caption{Fit 2a (12 branching ratios and 4 parameters): comparison of experimental with theoretical branching ratios. The fit parameters are found in 
Table~\ref{12b4ppa} and a best fit yields $\chi^2/\mathrm{d.o.f.}=12.73/(12-4)=1.59$ with DR and 
$\chi^2/\mathrm{d.o.f.}=10.95/(12-4)=1.37$ with CLFD.}
\label{12b4pbr}
\end{table}

\begin{table}
\begin{center}
\begin{tabular}{|c|ccc||cc|}  
\hline
   & $\nu_n$ &  $\theta_{\mathrm{mix}}$ &  $\rho_a$ &  $\nu_s$ & $N_S$ \\
\hline  \hline
\;\;CLFD \;\;   &\;\; $(3.19 \pm 0.32)\times 10^{-3}$\;\;&  $ 31.2^\circ \pm 3.7^\circ$ & \;\; $ 0.23\pm 0.05 $ \;\;&   \;\;   $(4.62 \pm 0.46)\times 10^{-3}$ \;\;&  \;\; 2.00 \;\; \\
\hline \hline
    DR    & $0.014\pm 0.012$  &    $40.9^\circ \pm 7.4^\circ$ &  $ 0.35\pm 0.42 $& $0.021\pm 0.017$  & 3.42   \\
\hline
\end{tabular}
\end{center}
\caption{Same as in Table~\ref{12b2ppa} but annihilation included (fit 2a, see Table~\ref{12b4pbr}). }
\label{12b4ppa}
\end{table}

\begin{table}
\begin{center}
\begin{tabular}{|c|c|c|c|c|c|}  
\hline
  Channel &  $\mathcal{BR}$ Exp.  & \;\; $\mathcal{BR}$ Th. (DR)\;\; &\;\; $\chi^2$\;\; &\;\; $\mathcal{BR}$ Th. (CLFD)\;\; & \;\; $\chi^2$ \;\; \\
\hline\hline
\;\;  $D^+ \to f_0 \pi^+$\;\; &&&&&\\
  (E791 [22]) &\;\; $(3.80\pm 1.17)\times 10^{-4}$\;\; & $2.63 \times 10^{-4}$ & 0.99 & $2.68 \times 10^{-4}$& 0.91\\
\hline 
\;\;  $D^0 \to f_0 \bar{K}^0$\;\; &&&&&\\
  (ARGUS [19])& \;\;$(6.40\pm 2.07)\times 10^{-3}$ \;\;&  $5.57 \times 10^{-3}$ & 0.15 & $5.57 \times 10^{-3}$&0.16\\
  (CLEO [17])& \;\;$(5.00\pm 1.52)\times 10^{-3}$ \;\;&  $5.57 \times 10^{-3}$ & 0.14 & $5.57 \times 10^{-3}$&0.14\\
  (BABAR [20]) & \;\;$(9.60\pm 8.55)\times 10^{-3}$ \;\;&  $5.57 \times 10^{-3}$ & 0.22 & $5.57 \times 10^{-3}$& 0.22\\
\hline
\;\;  $D^+ \to f_0 K^+$\;\; &&&&&\\
  (FOCUS [18])& \;\;$(1.22\pm 0.75)\times 10^{-4}$ \;\;& $1.42 \times 10^{-5}$ & 2.04 & $3.03 \times 10^{-5}$&1.48 \\
\hline
\;\;  $D_s^+ \to f_0 \pi^+$\;\; &&&&&\\
   (E791 [21])& $(1.14\pm 0.38)\times 10^{-2}$\;\;&  $1.37 \times 10^{-2}$ & 0.37 & $1.35 \times 10^{-2}$&0.32 \\ 
   (FOCUS [18])& \;\;$(1.90\pm 0.61)\times 10^{-2}$\;\;&  $1.37 \times 10^{-2}$ & 0.71 & $1.35 \times 10^{-2}$&0.76 \\
\hline
\;\;  $D_s^+ \to f_0 K^+$\;\; &&&&&\\
  (FOCUS [18])&\;\; $(2.24\pm 1.49)\times 10^{-3}$ \;\;& $0.75 \times 10^{-3}$ & 0.98 & $2.05 \times 10^{-3}$&0.01\\
\hline
\;\;  $D^0 \to f_0 \pi^0$ \;\;&&&&&\\
  (CLEO [46,47])&\;\; $(1.10\pm 0.97)\times 10^{-6}$ \;\;& $2.15\times 10^{-6}$ & 1.18 & $2.22 \times 10^{-6}$&1.35\\
\hline
\end{tabular}
\end{center}
\caption{Fit 2b (9 branching ratios and 4 parameters): comparison of experimental with theoretical branching ratios. The fit parameters 
are found in Table~\ref{9b4ppa} and a best fit yields $\chi^2/\mathrm{d.o.f.}=6.82/(9-4)=1.36$ with DR  and $\chi^2/\mathrm{d.o.f.}=5.37/(9-4)=1.07$ 
with CLFD.}
\label{9b4pbr}
\end{table}
\begin{table}
\begin{center}
\begin{tabular}{|c|ccc||cc|}  
\hline
   & $\nu_n$ &  $\theta_{\mathrm{mix}}$ &$\rho_a$ &       $\nu_s$ & $N_S$\\
\hline  \hline
 \;\;CLFD   \;\;  & \;\; $(3.09 \pm 0.36)\times 10^{-3}$ \;\;&  $ 31.5^\circ \pm 5.0^\circ$  &  \;\; $ 0.23\pm 0.25 $   \;\;  &    \;\;    $(4.49 \pm 0.52)\times 10^{-3}$  \;\;& \;\; 1.93 \;\; \\
\hline \hline
    DR    & $0.017\pm 0.010$  & $41.6^\circ \pm 7.1^\circ$ &    $ 0.34\pm 0.41 $       &      $0.024\pm 0.014$  & 3.84  \\
\hline
\end{tabular}
\end{center}
\caption{Same as in Table~\ref{12b4ppa} but for   fit 2b (see Table~\ref{9b4pbr}). }
\label{9b4ppa}
\end{table}

{\renewcommand\baselinestretch{1.40}
\begin{table}
\begin{center}
\begin{tabular}{|c|cc|cc|cc|}
\hline 
 $q^2$  & \multicolumn{2}{c}{$m_\pi^2 $}  & \multicolumn{2}{c}{$m_K^2$} & \multicolumn{2}{c|}{$m_\rho^2$}\\
 \hline \hline
  & \;\;CLFD\;\; & \;\;DR \;\;&\;\; CLFD\;\; &\;\; DR \;\;&\;\; CLFD \;\;& \;\;DR\;\;   \\
\;\; $F_0^{D\to f_0}(q^2)$ \;\;  & 0.21 & 0.22  & 0.28 &   0.18 & 0.38  &  0.17  \\
 \hline  
\;\; $F_1^{D\to f_0}(q^2)$ \;\;  & 0.21  & 0.22  & 0.33 &  0.22   & 0.94 &  0.26 \\
 \hline
\;\; $F_0^{D_s\to f_0}(q^2)$  \;\;   & 0.45 & 0.46  & 0.67 &  0.41  & 1.02 &  0.32  \\
 \hline  
\;\; $F_1^{D_s\to f_0}(q^2)$   \;\;  & 0.45 & 0.46  & 0.75 & 0.48 & 1.86 &  0.53  \\
\hline
\end{tabular}
\end{center}
\caption{Effective scalar and vector form factors $F_0(q^2)$ and $F_1(q^2)$ 
[see Eqs.~(\ref{effecffn}) and  (\ref{effecffs})] for various typical timelike momentum transfers, 
$q^2$, in $D_{(s)}\to f_0(980)$ transitions in the CLFD and DR approaches, respectively (see Figs.~\ref{ffDutofou} and~\ref{ffDstofos}).
\label{valffDtofo}}
\end{table}
}

{\renewcommand\baselinestretch{1.40}
\begin{table}
\begin{center}
\begin{tabular}{|c|cc|cc|cc|cc|}
\hline 
 $q^2$  & \multicolumn{2}{c}{$m_\pi^2 $}  & \multicolumn{2}{c}{$m_K^2$} & \multicolumn{2}{c}{$m_\rho^2$} 
&  \multicolumn{2}{c|}{$m_D^2$}  \\
 \hline \hline
  &\;\; CLFD\;\; & \;\;DR\;\; & \;\;CLFD \;\;& \;\;DR\;\; &\;\; CLFD\;\; &\;\; DR\;\; &\;\; CLFD\;\; &\;\; DR\;\;  \\
\;\; $F_0^{B\to f_0}(q^2)$ \;\;  & 0.12  & 0.12 & 0.13  &   0.12  & 0.14 &  0.12 & 0.23 &  0.13 \\
 \hline 
\;\; $F_1^{B\to f_0}(q^2)$ \;\;  & 0.12   & 0.12  & 0.14 &  0.12  & 0.15 &  0.13 & 0.28 & 0.15 \\
 \hline
\;\; $F_0^{B_s\to f_0}(q^2)$ \;\;    & 0.40 &  0.29  &0.41  &  0.30  &0.47  &  0.30 & 0.74 &  0.29  \\
 \hline  
\;\; $F_1^{B_s\to f_0}(q^2)$ \;\;    & 0.40 &  0.29  & 0.43 & 0.30  & 0.51 &  0.31 & 0.89 & 0.35 \\
\hline
\end{tabular}
\end{center}
\caption{Same as in Table~\ref{valffDtofo} but for  $B_{(s)}\to f_0(980)$ transitions (see Figs.~\ref{ffButofou} and~\ref{ffBstofos}).
\label{valffBtofo}}
\end{table}
}

{\renewcommand\baselinestretch{1.40}
\begin{table}[ph]
\begin{center}
\begin{tabular}{|c|c|cccc|}  
\hline
&    &  \;\;\; $D$ \;\;\; & \;\;\;$D_s$ \;\;\;&\;\;\; $B$ \;\;\;&\;\;\; $B_s$ \;\;\; \\
 \hline \hline
\;\; CLFD\;\; &\;\; $N_P$  \;\; & \;\;\; 9.976 & \;\;\; 7.340 \;\;\;& \;\;\;   5.880 & \;\;\; 3.833\;\;\; \\
& $\nu$   & \;\;\;0.046 & \;\;\; 0.061\;\;\;& \;\;\; 0.049 & \;\;\; 0.059\;\;\;  \\
\hline\hline 
 DR & \;\; $N_P$ \;\;  & \;\;\;4.395 &\;\;\; 3.443 \;\;\;& \;\;\; 3.937 & \;\;\; 2.646 \;\;\; \\
&  $\nu$    &\;\;\; 0.043 & \;\;\; 0.057 \;\;\;&\;\;\; 0.049 & \;\;\; 0.057\;\;\;\\ 
\hline
\end{tabular}
\end{center}
\caption{The pseudoscalar-meson parameters, $N_P$ and $\nu$, in the CLFD and  DR approaches. The normalization $N_P$ is either calculated with
Eq.~(\ref{eqnormps}) or Eq.~(\ref{eqnormps2}). The wave-function range parameter $\nu$, which enters the theoretical evaluation of the decay constants in Eq.~(\ref{fpclfd}) 
and Eq.~(\ref{fpdr}), is fitted to reproduce the experimental values of the decay constants of Eqs.~(\ref{eq5.14}).}
\label{pseudopa}
\end{table}
}
\end{document}